\documentclass[10pt,twocolumn,aps,prl,superscriptaddress,longbibliography,reprint]{revtex4-2}
\usepackage{setspace}
\usepackage{graphicx}
\usepackage{float}
\usepackage{ulem}
\usepackage{dcolumn}
\usepackage{hyperref}
\usepackage{multirow}
\usepackage{xcolor}
\usepackage{color}
\usepackage{soul,xcolor}
\usepackage{lineno}
\usepackage{amsmath}

\begin{document}

\title{Interlayer-coupling-driven stabilization and superconductivity in bilayer CoTe$_2$}

\author{Wenping Chen}
 \affiliation{Department of Physics, Jinan University, Guangzhou 510632, China}

\author{Ziyun Zhang}%
 \affiliation{Department of Physics, Jinan University, Guangzhou 510632, China}

\author{Feipeng Zheng}%
 \email{fpzheng\_phy@email.jnu.edu.cn}
 \thanks{corresponding author}
 \affiliation{Department of Physics, Jinan University, Guangzhou 510632, China}

\date{\today}

\begin{abstract}
Interlayer coupling plays a critical role in van der Waals materials by governing lattice stability and emergent quantum phases, yet its impact on few-layer hexagonal CoTe$_2$ remains unclear. 
Here, using first-principles calculations, we systematically investigate monolayer and bilayer CoTe$_2$ with an emphasis on their electronic structures, lattice dynamics, and electron-phonon coupling, and elucidate the underlying mechanisms driven by interlayer interactions. 
Our results show that monolayer CoTe$_2$ exhibits pronounced dynamical instability at low temperatures, whereas interlayer coupling stabilizes the bilayer crystal structure and gives rise to phonon-mediated superconductivity with a predicted critical temperature of about $4.7$~K when spin-orbit coupling is included. 
The stabilization and superconductivity in bilayer CoTe$_2$ are primarily attributed to interlayer-coupling-induced Te-$p_z$ charge redistribution and the associated modification of the Fermi surface and electron-phonon coupling.
Finally, we discuss how spin-orbit coupling in bilayer CoTe$_2$ weakens the EPC and superconductivity. 
Our work clarifies how interlayer coupling can jointly tune structural stability and superconductivity in few-layer CoTe$_2$, providing insights for engineering quantum phases in layered transition-metal dichalcogenides.
\end{abstract}

\maketitle

\section{introduction}

Van der Waals (VDW) materials are layered systems in which adjacent layers are held together by VDW forces. 
They can be grown from monolayer (1L) to multilayers using techniques such as mechanical exfoliation, molecular beam epitaxy, and chemical vapor deposition. 
The interlayer coupling in VDW materials can strongly influence their physical properties, leading to pronounced layer-dependent behaviors.
The ability to continuously tune the layer number enables systematic control of interlayer coupling, offering a powerful platform to investigate layer-dependent electronic orders, band topology, and low-dimensional quantum phenomena.

For example, in 1$T$-VTe$_2$, the charge density wave (CDW) order evolves with thickness. 
Bulk VTe$_2$ undergoes a structural transition from the hexagonal 1$T$ phase to the monoclinic 1$T^{\prime\prime}$ phase, induced by a $3\times 1 \times 3$ CDW \citep{ohtani1984phase,duvjir2022fine,mitsuishi2020switching}. 
In contrast, monolayer (1L) VTe$_2$ exhibits coexisting $4\times 4$ and $2\sqrt{3}\times 2\sqrt{3}$ CDW orders \citep{chazarin2024spatially,wu2020orbital,liu2020multimorphism,wang2025unraveling}. 
When reduced to a 1L, 1$T$-PtTe$_2$ undergoes a Lifshitz transition from a metal to a semiconductor with a band gap of about 0.79 eV \citep{lin2020dimensionality}. 
Weakening the interlayer coupling through intercalation transforms bulk 2$H$-NbSe$_2$ into a weakly coupled quasi-two-dimensional system, allowing it to exhibit Ising superconductivity with a relatively high critical temperature comparable to that of its bulk counterpart \citep{sun2023high,zhang2022tailored}.
These results demonstrate that investigating interlayer coupling in VDW materials not only deepens our understanding of their physical properties but also provides effective strategies for property tuning.

CoTe$_2$ is another transition-metal dichalcogenide, which exhibits multiple structural phases, including an orthorhombic phase, and a layered hexagonal (in $1T$ structure) phase.
Previous studies have shown that the orthorhombic phase exhibits magnetic order, which is influenced by both thickness and crystal orientation, displaying strong magnetic anisotropy \citep{slathia2024thickness,siegfried2023cote2}.
In contrast, bulk hexagonal CoTe$_2$ does not possess  long-range magnetic order \citep{chakraborty2023observation,kindra2024magnetism,wang2020thickness}.
Recent angle-resolved photoemission spectroscopy measurements combined with first-principles calculations further identify it as a type-II Dirac semimetal \citep{chakraborty2023observation}.
For few-layer CoTe$_2$, substantial progress has been made in both synthesis and materials engineering: multiple studies have reported the growth of few-layer CoTe$_2$, as well as the creation of derivative phases Co$_x$Te$_y$ via defect engineering, doping, and intercalation \citep{zhao2020engineering,wang2020thickness,zhao2024two,quan2024spontaneous,wu2024atomically,liu2024emerging}. 
For instance, thickness-controlled CoTe$_2$ films with thicknesses exceeding 4~nm have been synthesized, exhibiting electrical conductivities superior to those of many other transition-metal chalcogenides~\citep{wang2020thickness}. 
Moreover, experiments show that adjusting the relative chemical potentials of Co and Te allows the formation of bilayer (2L) CoTe$_2$ and self-intercalated structures such as Co$_9$Te$_{16}$. 
The self-intercalated phase displays kagome-like patterns in scanning tunneling microscopy and has been predicted to host flat-band features~\citep{wu2024atomically}. 
First-principles calculations further suggest that magnetic order---absent in pristine CoTe$_2$---can emerge in these self-intercalated structures~\citep{wu2024atomically,zhao2020engineering}. 
Despite these advances, a key question remains unresolved: how does interlayer coupling govern the intrinsic properties of pristine hexagonal CoTe$_2$? Systematic investigations targeting this issue are still lacking.

In this work, we employ first-principles calculations to systematically investigate the lattice dynamics, electronic structure, and electron-phonon coupling (EPC) in 1L and 2L 1$T$‑CoTe$_2$ (hereafter CoTe$_2$). 
Our results show that the 1L is dynamically unstable, exhibiting pronounced imaginary phonon modes primarily originating from out-of-plane Te (Te$_z$) vibrations  and in-plane Co (Co$_{xy}$) vibrations. 
In contrast, when going from 1L to 2L, interlayer coupling induces a redistribution of Te $p_z$ electrons, which strongly affects the Fermi surface and effectively suppresses the dynamical instability.
Meanwhile, phonon-mediated superconductivity emerges in the 2L, with an estimated critical temperature ($T_{\text{c}}$) of approximately 4.7 K.
We further clarify the respective roles of interlayer coupling and spin-orbit coupling (SOC) and analyze the mechanisms through which they influence the EPC.

\section{COMPUTATIONAL METHOD}

Density-functional theory (DFT)  and density-functional perturbation theory calculations were performed with the PBE exchange-correlation functional \citep{pbe1}.
We employed three types of pseudopotentials in our calculations: the projector-augmented wave (PAW) method \citep{paw1}, optimized norm-conserving Vanderbilt (ONCV) pseudopotentials \cite{oncv1,oncv2,oncv3}, and the norm-conserving pseudopotentials recommended by the PseudoDojo project \citep{dojo1}.
Van der Waals interactions in the 2L CoTe$_2$ structures were treated using Grimme's DFT-D3 correction with zero damping \citep{DFT-D3}, because this scheme reproduces the lattice parameters of bulk CoTe$_2$ reasonably well, as discussed below.
These calculations employed a combination of QUANTUM ESPRESSO (QE)\citep{qe1} and the Vienna Ab initio Simulation Package (VASP) \citep{vasp1}.
The VASP package, in conjunction with Phonopy \citep{phonopy1,phonopy2}, was used to generate a series of structures displaced along the eigenvector of the 1L imaginary phonon mode and to compute their total energies, from which the corresponding potential-energy surfaces are obtained.
Furthermore, the VASP package is also used as a calculator to obtain anharmonic phonon calculations in finite temperatures as described later.
The remaining calculations were performed by the QE package.
To simulate the thin-film geometry and reduce interactions between periodic boundaries, a vacuum layer of approximately 15 \AA~is introduced.
The Kohn-Sham valence states are expanded in a plane-wave basis. 
For PAW pseudopotentials, the kinetic-energy cutoffs are set to 50 Ry for the wave functions and 500 Ry for the charge density. 
For norm-conserving pseudopotentials, the corresponding cutoffs are 90 and 360 Ry, respectively.
The crystal structures are visualized using the VESTA package \citep{vesta1}.
Structural optimizations are carried out until the Hellmann-Feynman force acting on each atom is less than $1\times10^{-3}$ eV/\AA.
The samplings of electronic and phonon momenta in Brillouin zones (BZs) were done with  $\boldsymbol{k}$- and  $\boldsymbol{q}$-mesh, associated with grid spacing of $2\pi \times 0.015$  and $2\pi \times 0.045$~\AA$^{-1}$, respectively, whereupon  EPC matrix elements $g^{\nu}_{mn}(\boldsymbol{k}, \boldsymbol{q})$ were calculated~\citep{epw1}, which quantify the scattering amplitude between electronic states ($\boldsymbol{k}$,$m$) and ($\boldsymbol{k}+\boldsymbol{q}$,$n$) via phonon modes ($\boldsymbol{q}$, $\nu$), with $m$($n$) and $\nu$ being electronic and phonon band indexes, respectively. 
Subsequently, the above quantities were interpolated~\citep{w901} onto much denser grids with grid spacing of  $2\pi \times 0.0018$ ($\boldsymbol{k}$-grid) and $2\pi \times 0.0036$~\AA$^{-1}$ ($\boldsymbol{q}$-grid). 
Based on the above grids, the EPC-related quantities, including generalized static electronic susceptibility ($\chi_{\boldsymbol{q}\nu}$), Eliashberg function ($\alpha^{2}F(\omega)$), and momentum-resolved EPC constants ($\lambda_{\boldsymbol{q}\nu}$), were calculated. 
The electron-momentum-resolved superconducting gaps on the Fermi surface at temperature $T$, denoted as $\Delta_{\boldsymbol{k}}(T)$, were determined by solving the anisotropic Migdal-Eliashberg equations on an imaginary axis and then analytically continued to the real axis using  Padé functions \citep{anisotropicEPC1}.
In solving the equations, the Kohn-Sham states within 200 meV around the Fermi level are included, and the Matsubara frequencies are cutoff at 0.3 eV.

Anharmonic phonon calculations  were performed using the stochastic self-consistent harmonic approximation (SSCHA) \citep{sscha1,sscha2,sscha3}, a nonperturbative method that accounts for anharmonicity arising from both thermal and quantum fluctuations.
To calculate the anharmonic phonon of the 1L in finite temperatures, we use  2000 configurations with $6 \times 6$ supercells  in each population, to obtain the converged free energy Hessian.
Machine learning potentials, developed using the deep potential molecular dynamics method \citep{dpmd1,dpmd2}, are employed to model atomic interactions at various temperatures, enabling more efficient computation of the configurations in each population. 
The training sets comprise DFT-calculated energies, forces, and external pressures for 1000 configurations generated by SSCHA in each temperature.
The DFT calculations are performed by the VASP package with the same precision as the electronic structure calculations described earlier.
To achieve optimal concordance between DFT-computed properties and those predicted by the machine learning potential, the loss function—encompassing energy, force, and external pressure contributions—undergoes mini-mization through $4 \times 10^{6}$ iterative optimization steps.
A comparative analysis of energies, forces, and external pressures derived from DFT calculations and machine learning potential predictions shown in Fig.~S1 \cite{SM}.

\section{results and discussions}

\subsection{Dynamical instability in 1L CoTe$_2$}

\begin{figure*}[htb]
\includegraphics{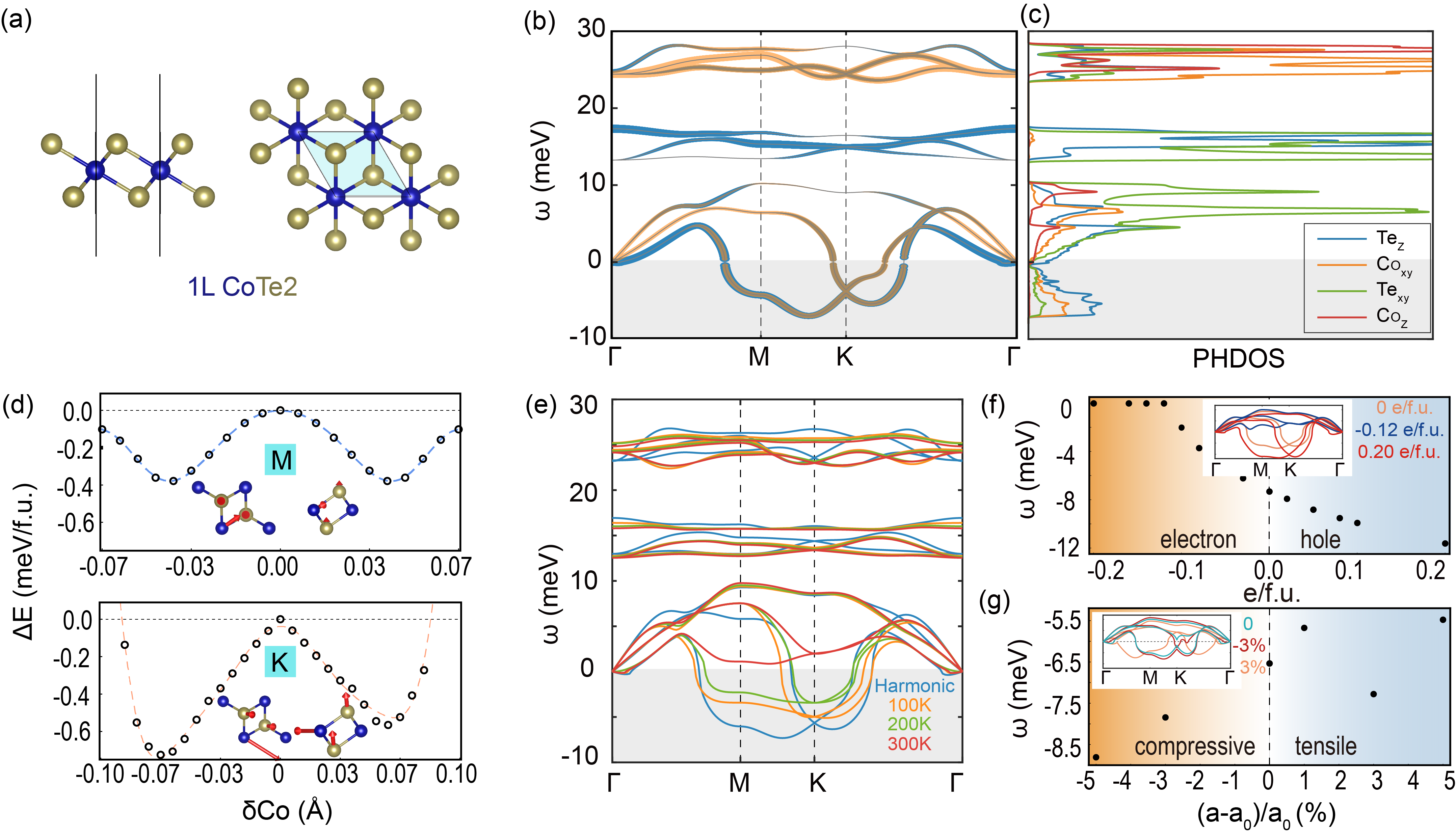}
\caption{\label{fig1} 
Lattice-dynamical properties of 1L CoTe$_2$:
(a) side and top views of its crystal structure;
(b) the $\omega_{\boldsymbol{q}\nu}$ calculated using an electron broadening of 0.01 Ry, and projected onto Te$_z$ and Co$_{xy}$ vibrations, where the blue (orange) symbol size is proportional to the Te$_z$ (Co$_{xy}$) contribution; 
(c) vibration-resolved phonon density of states; 
(d) potential-energy surfaces along the eigen-displacement of the imaginary mode at M (upper panel) and of one imaginary mode at K (lower panel). 
Insets in (d) display the corresponding phonon polarization vectors in top and side views; 
(e) harmonic $\omega_{\boldsymbol{q}\nu}$,  and anharmonic $\omega_{\boldsymbol{q}\nu}$ obtained from the SSCHA free-energy Hessian at selected temperatures; 
(f, g) the Minimum phonon frequency as a function of charge doping and biaxial strain, respectively; the insets show the acoustic-branch dispersions at three representative doping levels and three biaxial strains, respectively.
}
\end{figure*}

Bulk hexagonal CoTe$_2$ can be regarded as AA-stacked CoTe$_2$ 1Ls separated by an interlayer spacing, which is calculated to be 2.66~\AA~in this work. 
Within each 1L, each Co atom resides at the center of an octahedron coordinated by six neighboring Te atoms. 
Our calculated lattice parameters (a = 3.75~\AA, c = 5.45~\AA) for bulk CoTe$_2$ are in close agreement with a recent report (a = 3.80~\AA, c = 5.41~\AA)  \citep{chakraborty2023observation}. 
Furthermore, the electronic band structure we obtain for bulk CoTe$_2$ (Fig.~S2 \citep{SM}) is consistent with both the theoretical results and the angle-resolved photoemission spectroscopy measurements presented in the same study \citep{chakraborty2023observation}. 
These agreements support the soundness of our computational setup and strengthen confidence in our description of hexagonal CoTe$_2$. 
We next turn to the investigation of 1L CoTe$_2$.

Interestingly, our phonon calculations indicate that 1L CoTe$_2$ is dynamically unstable, in contrast to isostructural 1L NiTe$_2$ \citep{zheng2020emergent}, containing the neighboring 3d transition metal of Ni.
As shown in Fig.~\ref{fig1}(b), the phonon dispersion ($\omega_{\boldsymbol{q}\nu}$) exhibits imaginary frequencies near the M and K points, originating from two acoustic branches. 
The most unstable modes occur at the  midpoint of the M--K path (MK/2). 
According to the vibration-resolved phonon density of states (PHDOS) in Fig.~\ref{fig1}(c), these imaginary modes are dominated by out-of-plane vibrations of Te atoms (Te$_z$) and in-plane vibrations of Co atoms (Co$_{xy}$). 
This assignment is further corroborated by the real-space eigenvectors of the imaginary modes at the M and K points (Fig.~\ref{fig1}(d)), which clearly feature Te$_z$ and Co$_{xy}$ character. 
Displacing atoms along these eigenvectors yields double-well potential energy surfaces (PES) as shown in Fig.~\ref{fig1}(d), consistent with the presence of imaginary phonon frequencies. 
The $\omega_{\boldsymbol{q}\nu}$ computed using different pseudopotentials (see Fig.~S3 \citep{SM}) and those obtained with spin-orbit coupling (SOC) included (see Fig.~S4(a) \citep{SM}) show similar behavior, leading to the same conclusion.
Moreover, when anharmonic effects at finite temperature are included using the SSCHA, the dynamical instability is suppressed by thermal fluctuations. 
As shown in Fig.~\ref{fig1}(e), the imaginary modes progressively harden with increasing temperature and become entirely positive at 300 K. 
Taken together, these results suggest that 1L CoTe$_2$ is dynamically unstable at low temperature, with the instability primarily associated with Te$_z$ and Co$_{xy}$ vibrations.
Such an intrinsic instability may help rationalize the experimentally reported difficulty in stabilizing 1L CoTe$_2$ on SiO$_2$/Si \citep{wang2020thickness}.

We further examine how in-plane biaxial strain and charge doping affect the dynamical stability. 
Fig.~\ref{fig1}(f) summarizes the dependence of the minimum phonon energy ($\omega_{\text{min}}$) on biaxial strain and carrier doping. 
Over the strain range from $-5\%$ (biaxial compression) to $+5\%$ (biaxial tension), $\omega_{\text{min}}$ remains below $-5.5~\mathrm{meV}$, indicating persistent dynamical instability. 
Regarding doping, hole doping drives $\omega_{\text{min}}$ to more negative values, thereby enhancing the instability, while electron doping increases $\omega_{\text{min}}$, progressively suppressing it. 
The variation of $\omega_{\text{min}}$ with doping can be rationalized, as the Fermi surface of the 1L consists of hole pockets, as we will show later.
The $\omega_{\text{min}}$ reaches zero only above an electron-doping level of approximately $-0.12~\mathrm{e/f.u.}$, at which point the system becomes dynamically stable. 
Overall, these calculations show that the dynamical instability of 1L CoTe$_2$ persists under moderate in-plane strain and charge doping, but is fully removed under sufficiently strong electron doping.

\begin{figure}[htb]
\includegraphics{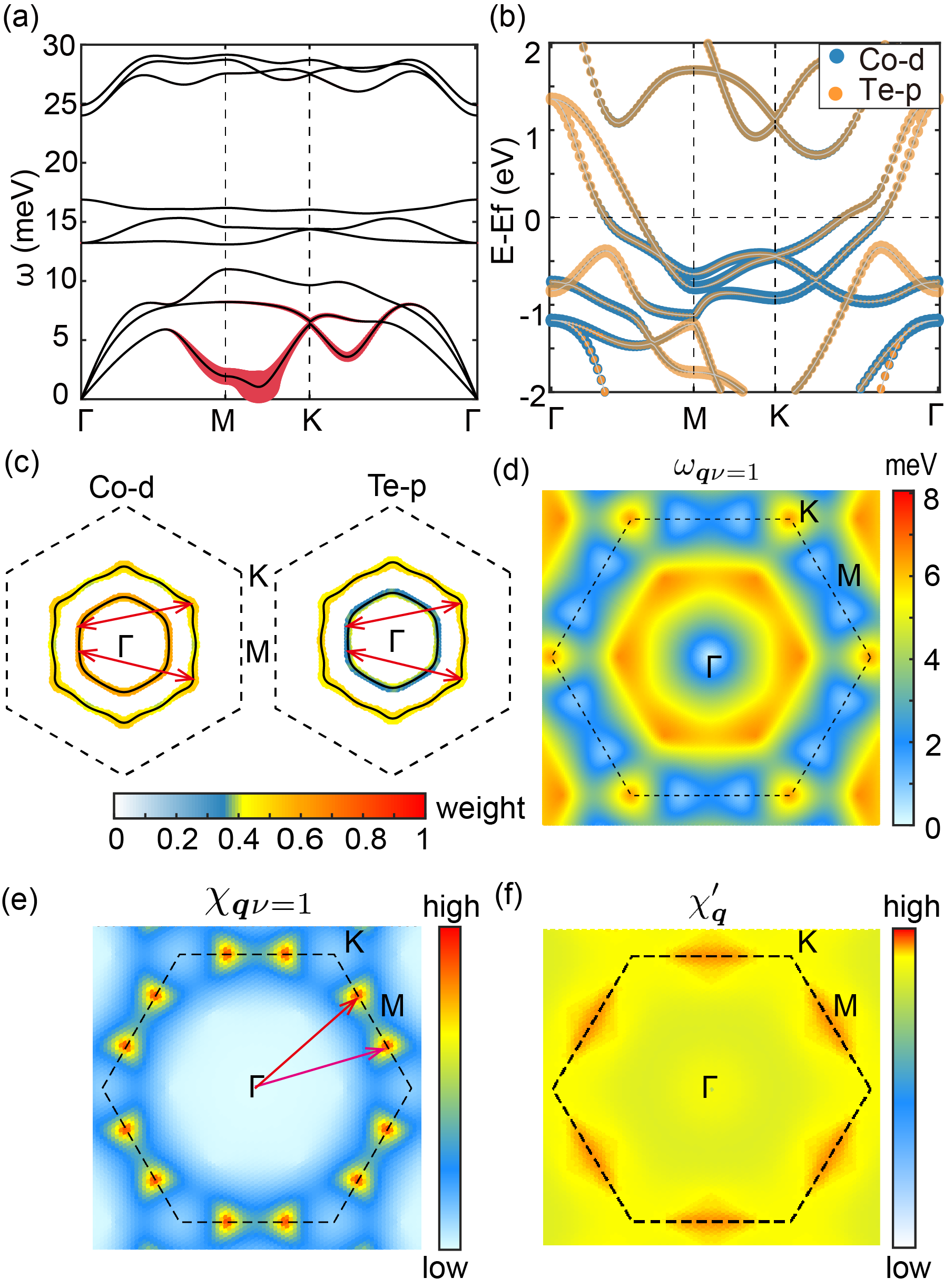}
\caption{\label{fig2} 
Calculated quantities elucidating the mechanism of phonon softening in 1L CoTe$_2$ are summarized in (a--f). 
(a) The $\omega_{\boldsymbol{q}\nu}$ computed with a relatively large electronic smearing (0.018 Ry); the red symbol size scales with $\lambda_{\boldsymbol{q}\nu}$. 
(b) Electronic band structure with orbital projections onto Co-$d$ and Te-$p$ states.
(c) Fermi surfaces projected onto Co-$d$ and Te-$p$ states; double-headed arrows mark the $\boldsymbol{q}$ vectors where $\chi_{\boldsymbol{q},\nu=1}$ is strongly enhanced (see panel (e)). 
(d) Map of $\omega_{\boldsymbol{q}\nu=1}$ across an extended BZ. 
(e) Generalized static electronic susceptibility $\chi_{\boldsymbol{q}\nu=1}$. 
(f) Static susceptibility within the constant-matrix approximation, $\chi^{\prime}_{\boldsymbol{q}}$.
}
\end{figure}

To gain deeper insight into the microscopic origin of the dynamical instability in the 1L, we evaluate its electron-phonon coupling (EPC), since in many transition-metal compounds imaginary phonon frequencies are often associated with excessively strong EPC \citep{yan2025charge,wang2023decisive,luo2023emergent}.
Phonon softening induced by EPC at branch $\nu$ and wave-vector $\boldsymbol{q}$ can be quantified in terms of the generalized static electronic susceptibility \citep{giustino2017electron,felix2016,zhu2015classification}, which is directly related to the real part of the EPC-induced phonon self-energy:
$$
\chi_{\boldsymbol{q}\nu}=\sum_{\boldsymbol{k},m,n}w_{\boldsymbol{k}}|g^{\nu}_{m\boldsymbol{k},n\boldsymbol{k}+\boldsymbol{q}}|^2\frac{f(\varepsilon_{m\boldsymbol{k}})-f(\varepsilon_{n\boldsymbol{k}+\boldsymbol{q}})}{\varepsilon_{n\boldsymbol{k}+\boldsymbol{q}}-\varepsilon_{m\boldsymbol{k}}},
$$
where $w_{\boldsymbol{k}}$ is the weight of the $\boldsymbol{k}$-point for BZ integration, $f(\varepsilon_{m\boldsymbol{k}})$ is the Fermi-Dirac distribution function evaluated at the electronic energy $\varepsilon_{m\boldsymbol{k}}$ associated with Kohn-Sham state ($m,\boldsymbol{k}$). 
Within the constant-matrix-element approximation, the $\chi_{\boldsymbol{q}\nu}$~reduces to the bare electronic susceptibility $\chi^{\prime}_{\boldsymbol{q}}$:
$$
	\chi^{\prime}_{\boldsymbol{q}}=\sum_{\boldsymbol{k}, m, n} w_{\boldsymbol{k}} \frac{f\left(\varepsilon_{m\boldsymbol{k}}\right)-f\left(\varepsilon_{n\boldsymbol{k}+\boldsymbol{q}}\right)}{\varepsilon_{n\boldsymbol{k}+\boldsymbol{q}} - \varepsilon_{m\boldsymbol{k}}},
	\label{eqCHIPQ}
$$
which captures the contribution from Fermi-surface nesting, i.e., a purely electronic effect.
As the EPC matrix elements $g^{\nu}_{m\boldsymbol{k},n\boldsymbol{k}+\boldsymbol{q}}$ are only well defined when $\omega_{\boldsymbol{q}\nu}>0$, we recomputed the $\omega_{\boldsymbol{q}\nu}$ using a large electronic broadening of 0.018 Ry to remove the imaginary modes \citep{zheng2019electron}. 
As shown in Fig.~\ref{fig2}(a), the recalculated $\omega_{\boldsymbol{q}\nu}$ display pronounced softening near the M and K points, with the most pronounced dip occurring along the M--K path. 
This is more clearly visualized in Fig.~\ref{fig2}(d), which plots the distribution of the lowest branch, $\omega_{\boldsymbol{q}\nu=1}$, in an extended BZ.
The above feature is consistent with the $\omega_{\boldsymbol{q\nu}}$ obtained using a regular electronic smearing (Fig.~\ref{fig1}(b)). 
Based on the recalculated $\omega_{\boldsymbol{q}\nu}$,  we evaluated $\chi_{\boldsymbol{q}\nu=1}$ and $\chi'_{\boldsymbol{q}}$, as shown  in Figs.~\ref{fig2}(e) and \ref{fig2}(f), respectively.
A direct comparison shows that $\chi'_{\boldsymbol{q}}$  (Fig.~\ref{fig2}(f)) is strongly enhanced in broad regions centered at the M points, including MK/2 and its symmetry-related wave vectors, where the phonon softening is most pronounced.
After including the EPC matrix elements $g^{\nu}_{m\boldsymbol{k},n\boldsymbol{k}+\boldsymbol{q}}$, these broad high-$\chi^{\prime}$ regions collapse into isolated hotspots along M--K paths, coincided with the $\boldsymbol{q}$ vectors where the strongest phonon softening occurs (Fig.~\ref{fig2}(d)).
These results indicate that the phonon softening in the 1L arises from the combined effect of Fermi-surface nesting and momentum-dependent EPC matrix elements: the former enhances the relevant $\boldsymbol{k} \rightarrow \boldsymbol{k}+\boldsymbol{q}$ scattering channels, while the latter further weights these channels and localizes the strongest instability at specific wave vectors around MK/2.

The pronounced EPC can be traced to inter-pocket scattering between the $p$--$d$-hybridized electronic states. 
Fig.~\ref{fig2}(b) shows the orbital-projected band structure of the 1L, resolved into $d$ orbital states of Co atoms (Co-$d$) and $p$ orbital states of Te atoms (Te-$p$).
The calculated band structure is consistent with previous studies \citep{wang2020thickness,liu2024emerging,zhao2024two}.
One can see in Fig.~\ref{fig2}(b) that along the $\Gamma$--M and $\Gamma$--K directions, two bands cross the Fermi level, giving rise to two hole pockets centered at $\Gamma$. 
The energy states near the Fermi surface are dominated by hybridized Co-$d$ and Te-$p$ character. 
More specifically, the inner hole pocket has a larger Co-$d$ weight, whereas the outer pocket has comparable Co-$d$ and Te-$p$ weights (Fig.~\ref{fig2}(c)).
We find that these two pockets are mutually nested by $\boldsymbol{q}$ vectors, of which $\chi_{\boldsymbol{q}\nu=1}$ is large.
In Fig.~\ref{fig2}(c), two representative nesting vectors are indicated by red double-headed arrows, which connect substantial portions of the inner and outer pockets.
These two nesting vectors  exhibit pronounced $\chi_{\boldsymbol{q}\nu=1}$, as shown in Fig.~\ref{fig2}(e).
The above results indicate that the dynamical instability in the 1L is primarily driven by the enhanced EPC near the wave vectors of MK/2, arising from inter-pocket scattering, which is associated with $p$--$d$-hybridized electronic states.
As including SOC does not significantly alter the electronic structure near the Fermi level (see Fig.~S4(b) \citep{SM}), the above conclusions are expected to remain unchanged.

\subsection{Dynamical stability of 2L CoTe$_2$}

\begin{figure*}[htb]
\includegraphics{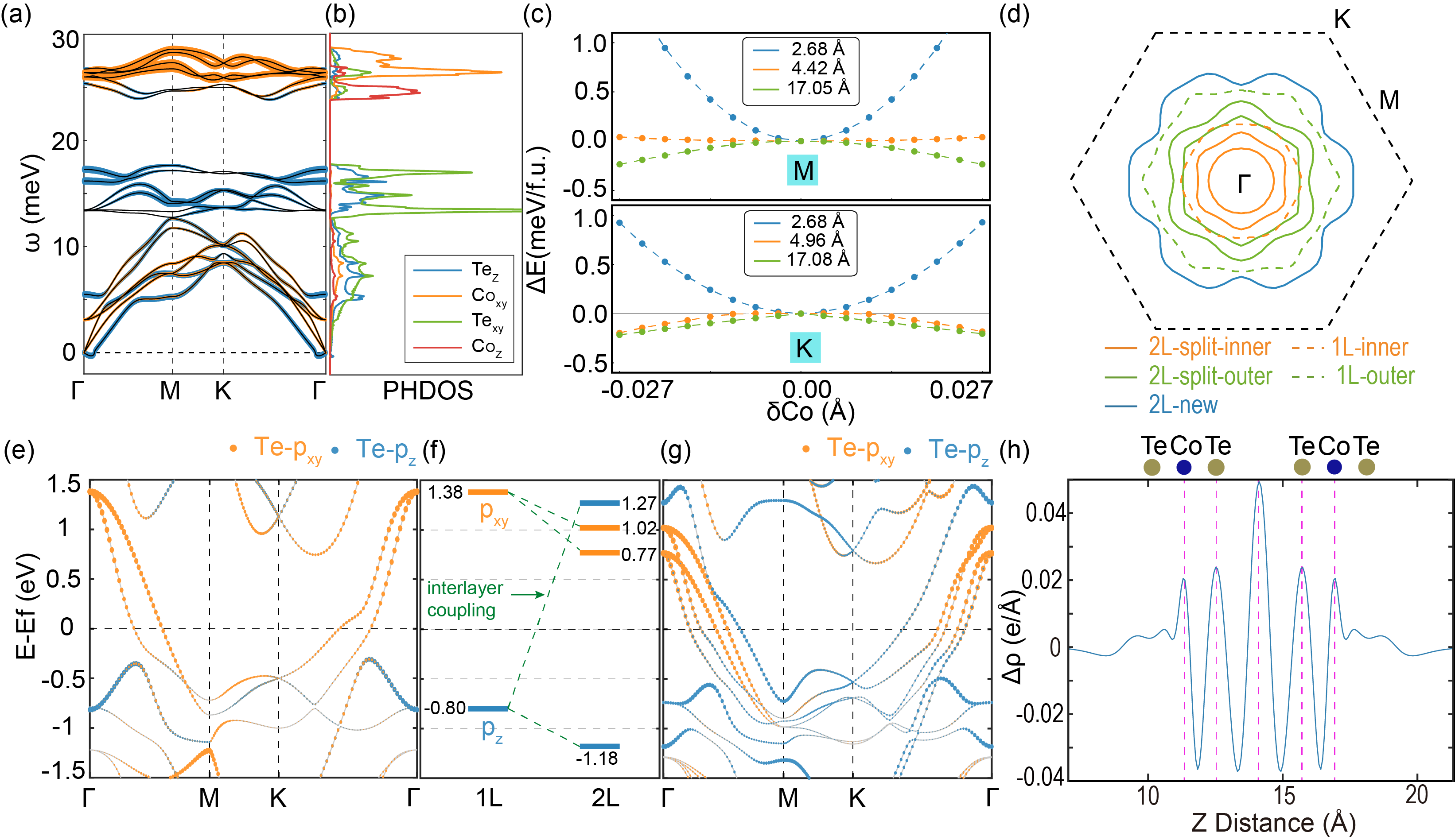}
\caption{\label{fig3} 
 Calculated quantities elucidating effect of interlayer coupling in 2L CoTe$_2$ are summarized in (a--h).
 (a) The $\omega_{\boldsymbol{q}\nu}$ projected onto Te$_z$ and Co$_{xy}$ vibrations; the blue (orange) symbol size scales with the Te$_z$ (Co$_{xy}$) contribution.
 (b) Vibration-resolved phonon density of states; 
(c) Potential-energy surfaces for an AA-stacked bilayer constructed from two 1L CoTe$_2$ sheets at different interlayer separations, with atomic displacements along the eigenvectors of the 1L imaginary mode at M (upper) and K (lower) (Fig.~\ref{fig1}(d)). 
 (d) Fermi surfaces of the 2L (solid) and 1L (dashed) CoTe$_2$.
 (e,g) Electronic band structures of 1L and 2L, respectively, projected onto the Te-$p_{xy}$ and  Te-$p_{z}$ states.
 (f) Schematic of the interlayer-coupling-induced band splitting at the $\Gamma$ when going from a 1L to a 2L CoTe$_2$.
 (h) Planar integral (along $xy$)  of charge-density difference for 2L CoTe$_2$ relative to two isolated 1Ls, ($\Delta \rho(z)$), which quantifies net charge accumulation/depletion per $z$ plane along the stacking direction. 
 See Fig.~S6 \citep{SM} for more details.
}
\end{figure*}

Interestingly, we find that the dynamical instability is suppressed when going from the 1L to the 2L structure.
Our calculations show that, for 2L CoTe$_2$, the AA-stacked 1T structure is the most energetically favorable among the stacking configurations considered (see Fig.$~$S5 \citep{SM}), consistent with its bulk counterpart.
Fig.~\ref{fig3}(a) shows the $\omega_{\boldsymbol{q}\nu}$ of the 2L CoTe$_2$. 
With the exception of a very small region of $\boldsymbol{q}$ near the $\Gamma$ point, in which a few imaginary modes emerge (the lowest energy about $-0.29~\mathrm{meV}$), all  $\omega_{\boldsymbol{q}\nu}$ are positive. 
The weak imaginary modes are attributable to numerical noise and are therefore negligible.
By comparing the  $\omega_{\boldsymbol{q}\nu}$  and the PHDOS of the 1L (Figs.~\ref{fig1}(b) and \ref{fig1}(c)) with those of the 2L (Figs.~\ref{fig3}(a) and \ref{fig3}(b)), it is evident that the imaginary phonon modes in the two lowest branches---associated with Te$_z$ and Co$_{xy}$ vibrations---are fully eliminated in the 2L. 
These instabilities evolve into phonon modes that exhibit locally flattened dispersions in the vicinity of the  M and
K points, with characteristic energies of about 5.1, 7.3, and 8.4~meV, thereby producing the three corresponding PHDOS peaks in Fig.~\ref{fig3}(b).
These modes are relevant to the superconductivity in the 2L, as discussed later.

The absence of imaginary phonon modes in the 2L indicates that interlayer coupling is crucial for stabilizing the structure.
To verify this conjecture, we constructed 2L structures with different interlayer spacings and calculated the potential energy surfaces (PESs) along the atomic displacement directions, corresponding to the imaginary modes at the M and K points of the 1L (insets of Fig.~\ref{fig1}(d)).
As shown in Fig.~\ref{fig3}(c), when the interlayer spacing exceeds 17~\AA, the interlayer coupling is weak, and the PESs at both the M and K points exhibit a concave shape, consistent with the 1L results. 
As the interlayer spacing decreases toward the fully relaxed bilayer distance (2.66~\AA), the PES gradually evolves from concave to convex, corresponding to positive vibrational frequencies. 
These results show that interlayer coupling is essential for stabilizing the 2L structure. 
The underlying mechanism is analyzed below from an electronic-structure perspective.

We first examine how the electronic structure evolves from the 1L to the 2L.
Figs.~\ref{fig3}(e) and~\ref{fig3}(g) show the band structures of the 1L and 2L, respectively. 
The most prominent change is that the two bands crossing the Fermi level in the 1L split into pairs in the 2L  due to interlayer coupling. 
Furthermore, a new band appears in the 2L. 
It disperses along the $\Gamma$--M direction, crosses the Fermi level near the M point, extends up to about 1.5~eV, and forms a band maximum near the $\Gamma$ point. 
Fig.~\ref{fig3}(f) illustrates this band splitting behavior using the electronic states at the $\Gamma$ point as an example. 
These evolutions in the electronic structure indicate that a Lifshitz transition \citep{lifshitz1960anomalies} occurs in the 2L due to interlayer coupling.
The band splitting leads to a substantial reconstruction of the Fermi surface. 
As shown in Fig.~\ref{fig3}(d), five Fermi pockets are present in the 2L.
Based on the corresponding band structure in Fig.~\ref{fig3}(g), they are identified as hole pockets.
More specifically, the outermost petal-shaped pockets pointing toward the K and K$^\prime$ points originate from the new pockets induced by the Lifshitz transition, while the remaining four pockets result from the splitting of the two hole pockets in the 1L. 
Furthermore, the split hole pockets are significantly smaller than the original hole pockets in the 1L (see Fig.~\ref{fig3}(d)).
Overall, the most significant changes in the 2L are the emergence of an additional hole pocket associated with the Lifshitz transition, and the splitting and shrinkage of the original 1L hole pockets.

These electronic structure changes arise from charge redistribution induced by interlayer coupling. 
When two 1Ls approach to form a 2L system, the electronic density undergoes pronounced reconstruction. 
As shown in Fig.~\ref{fig3}(h), the electron density associated with the out-of-plane $p_z$ orbitals of Te atoms (Te-$p_z$) on both sides of the interlayer region is significantly reduced. 
Concurrently, there is a notable enhancement of electron density both in the middle of the van der Waals gap (along the Te-Te bonding direction) and within the in-plane directions of the Te atoms (see Fig.~S6 for more details \citep{SM}).
This indicates that these Te-$p_z$ electrons from different layers hybridize to form interlayer (quasi) bonds, accompanied by partial charge transfer to in-plane orbitals. 
This electronic reconstruction naturally explains the pronounced splitting of $p_z$-orbital-related bands and the occurrence of the Lifshitz transition in the 2L: as the Te-$p_z$ orbitals evolve from fully occupied to partially occupied, the original hole pockets effectively experience electron doping and shrink.

Based on these results, the mechanism of the stabilization of the 2L crystal structure can be understood as follows.
On the one hand, the formation of interlayer Te--Te bonds strengthens the force constants associated with Te$_z$ vibrations, thereby hardening the corresponding phonon modes.
On the other hand, interlayer-coupling-induced splitting and shrinkage of the hole pockets suppress the EPC strength near the wave vector MK/2 and its symmetry-equivalent points, which in turn effectively stabilizes the 2L crystal structure.
The calculated real part of the phonon self-energy for the 1L and 2L systems (Fig.~S7 \citep{SM}) further supports these conclusions.
The stabilization of the 2L is consistent with experimental results that the resistivity of few-layer CoTe$_2$ exhibits no abrupt anomaly from 300 down to 2~K \cite{wang2020thickness}.
In addition to interlayer coupling, self-intercalation  in experimentally realized Co-Te compounds may further contribute to the stabilization of few-layer structures by enhancing interlayer bonding.

\subsection{Emergent superconductivity in 2L CoTe$_2$ and the SOC effect}

\begin{figure}[htb]
\includegraphics{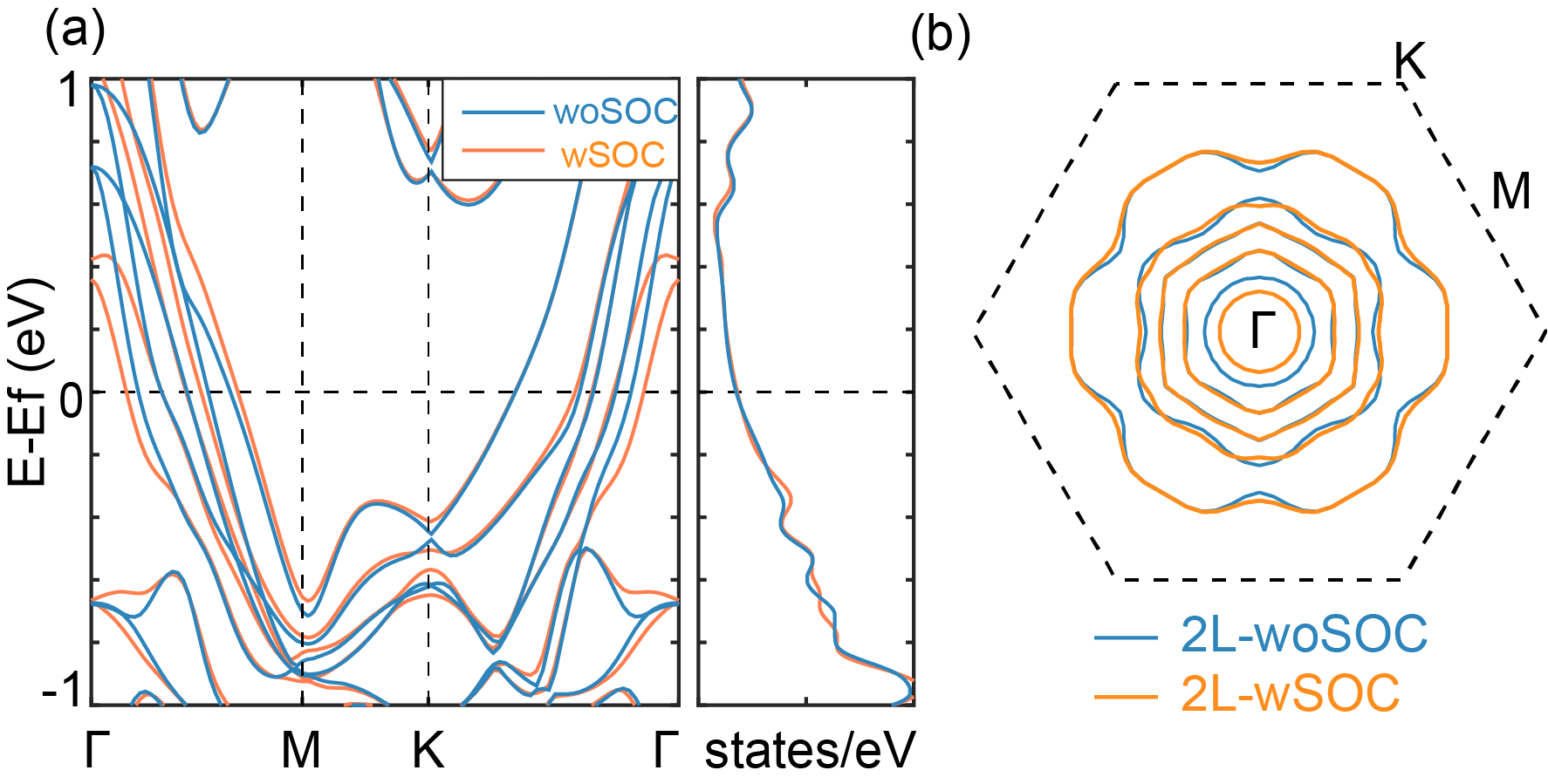}
\caption{\label{fig4} 
 (a) Band structures, density of states, and (b) Fermi surfaces of the 2L, calculated with SOC (wSOC) and without SOC (woSOC).
}
\end{figure}

As discussed above, the imaginary modes in the 1L evolve into low-energy modes with flattened dispersions in the 2L. 
This prompts an investigation of their superconducting potential.
We first examine effect of SOC on the electronic structure. 
Figs.~\ref{fig4}(a) and~4(b) show the calculated band structures and Fermi surfaces with and without SOC, respectively. 
As the system preserves inversion symmetry, SOC does not lift the spin degeneracy of the bands but instead shifts their energies. 
A comparison of the Fermi surfaces (Fig.~\ref{fig4}(b)) reveals that the most pronounced effect is a contraction of the hole pocket closest to $\Gamma$ point. 
This is also evident in the band structure along the $\Gamma$--M and $\Gamma$--K directions in Fig.~\ref{fig4}(a). 
Meanwhile, the outer hole pocket undergoes a slight expansion along $\Gamma$--M, compensating for the loss of states from the shrinking $\Gamma$-centered pocket. 
This electronic redistribution results in a nearly unchanged density of states at the Fermi level, $N(0)$, upon including SOC (right panel of Fig.~\ref{fig4}(a)).
Nevertheless, as will be discussed, the superconductivity is weakened due to SOC, primarily by the contraction of the $\Gamma$-centered hole pocket.

\begin{figure*}[htb]
\includegraphics{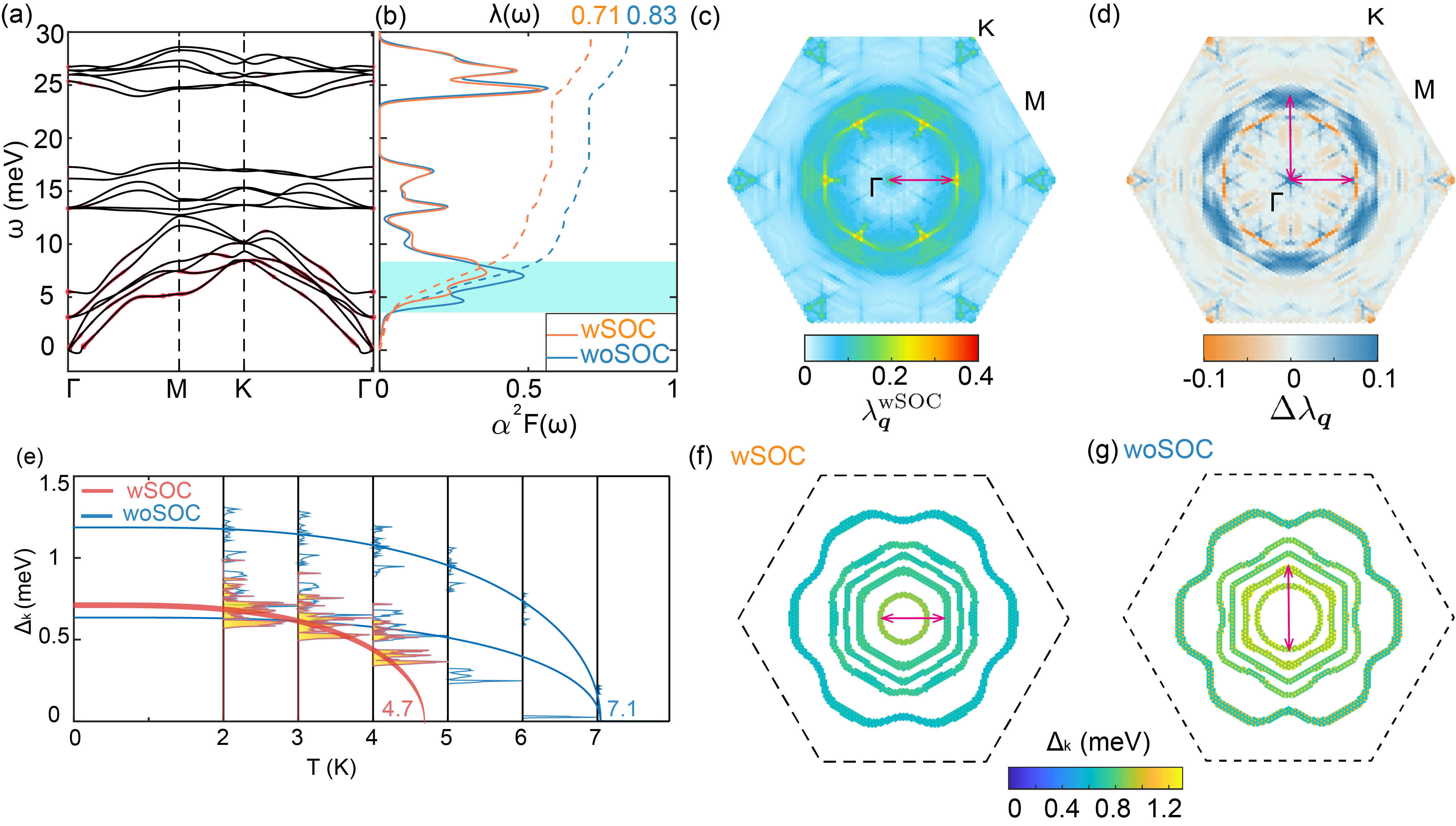}
\caption{\label{fig5} 
  Calculated quantities elucidating the EPC, the superconducting mechanism, and the SOC effect in the 2L.
  (a) The $\omega_{\boldsymbol{q}\nu}$  computed using an electron broadening of 0.01 Ry; the red symbol size scales with $\lambda_{\boldsymbol{q}\nu}$. 
  (b) $\alpha^{2}F(\omega)$ and $\lambda(\omega)$ calculated with and without SOC;
  the shaded region marks where SOC induces the largest deviations.
  (c) BZ distribution of $\lambda^{\text{SOC}}_{\boldsymbol{q}}(\omega_1,\omega_2)$ with $\omega_1 = 3.50$ meV and $\omega_2 = 8.15$ meV, quantifying the contribution of the phonon modes $(\boldsymbol{q},\nu)$ within the energy window $[\omega_1,\omega_2]$ to $\alpha^{2}F(\omega)$.
  (d) BZ distribution of $\Delta\lambda_{\boldsymbol{q}}(\omega_1,\omega_2)=\lambda^{\text{woSOC}}_{\boldsymbol{q}}(\omega_1,\omega_2)- \lambda^{\text{SOC}}_{\boldsymbol{q}}(\omega_1,\omega_2)$.
  The double-headed arrows in (c) and (d) denote representative $\boldsymbol{q}$ vectors where $\lambda^{\mathrm{SOC}}_{\boldsymbol{q}}(\omega_1,\omega_2)$ and $|\Delta\lambda_{\boldsymbol{q}}|$ are particularly pronounced, respectively.
  (e) Histograms of temperature-dependent superconducting gaps $\Delta_{\boldsymbol{k}}(T)$, for those electronic states, whose Kohn-Sham energies fall within $\pm 0.1$ eV of the Fermi level.
  The colored curves show the BCS fits to the gaps calculated with and without SOC.
  (f, g) BZ distribution of $\Delta_{\boldsymbol{k}}(T)$ at $T = 2$ K calculated with and without SOC, respectively;
  the double-headed arrows are reproduced from panel (d).
}
\end{figure*}

Furthermore, we evaluate the superconducting properties of the 2L arising from EPC. 
The orange solid and dashed curves in Fig.~\ref{fig5}(b) present the Eliashberg function $\alpha^{2}F(\omega)$ and the cumulative EPC strength $\lambda(\omega)$ calculated with SOC included. 
The total EPC constant ($\lambda$) is 0.71, with the dominant contributions originating from three phonon energy windows: 0--12.3~meV, 12.3--18.2~meV, and 23.5--29.2~meV, contributing 0.50, 0.08, and 0.13 to the $\lambda$, respectively. 
Notably, the low-energy phonons provide the largest contribution, accounting for 70\% of the $\lambda$. 
This behavior is also evident from the momentum-resolved distribution of $\lambda_{\boldsymbol{q}\nu}$ shown in Fig.~\ref{fig5}(a). 
To gain further insight into the microscopic origin of superconductivity, we plot in Fig.~\ref{fig5}(c) the contribution of phonons in an energy range 3.50--8.15~meV (highlighted in Fig.~\ref{fig5}(b)) to $\lambda_{\boldsymbol{q}}$, written as:
$$
\lambda^{\text{wSOC}}_{\boldsymbol{q}}(\omega_1,\omega_2) = \sum_{\nu}\int_{\omega_1}^{\omega_2}\mathrm{d}\omega \lambda_{\boldsymbol{q}\nu}\delta(\omega - \omega_{\boldsymbol{q}\nu}),
$$
where $\omega_1=3.50$~meV and  $\omega_2=8.15$~meV, respectively.
It can be seen in Fig.~\ref{fig5}(c) that the $\boldsymbol{q}$ of these phonons are concentrated around a hexagonal contour centered at the $\Gamma$ point, with its six vertices oriented toward the M points. 
A representative $\boldsymbol{q}$ exhibiting a large $\lambda_{\boldsymbol{q}}$, which lies at the midpoint of one side of this hexagon, is marked with a double-headed arrow.
Subsequent analysis reveals that these phonons predominantly mediate the coupling between electronic states on the two hole pockets closest to the $\Gamma$ point, as discussed below.

To quantitatively assess the superconducting properties, we further take into account the anisotropy of the EPC. 
We calculate the momentum-resolved superconducting gap $\Delta_{\boldsymbol{k}}(T)$, which denotes the superconducting pairing amplitude for electrons at momentum $\boldsymbol{k}$ and temperature $T$, for states near the Fermi level at various temperatures.
The statistical histograms of $\Delta_{\boldsymbol{k}}(T)$ are shown in Fig.~\ref{fig5}(e).
One can see that at $T = 2$~K, $\Delta_{\boldsymbol{k}}$ spans the range of 0.5--1.0~meV with SOC included. 
More specifically, a small peak appears around 1.0~meV, while the majority of $\Delta_{\boldsymbol{k}}$ values are concentrated between 0.5 and 0.8~meV. 
By correlating these features with Fig.~\ref{fig5}(f), which  displays the distribution of these $\Delta_{\boldsymbol{k}}$ in the BZ, we find that the higher-$\Delta_{\boldsymbol{k}}$ peak at approximately 1.0~meV mainly originates from the hole pocket closest to the $\Gamma$ point, whereas the $\Delta_{\boldsymbol{k}}$ in the range of 0.5--0.8~meV are assigned to the remaining four hole pockets. 
A direct comparison between Fig.~\ref{fig5}(f) and the double-headed arrows in Fig.~\ref{fig5}(c) further confirms that the electronic states dominating the EPC are primarily located on the two innermost hole pockets, indicating that the EPC in 2L CoTe$_2$ mainly originates from inter-pocket scattering between these two hole pockets.
Furthermore, as the temperature increases, $\Delta_{\boldsymbol{k}}$ decreases continuously and eventually vanishes at around 4.7~K, indicating a $T_{\text{c}}$ of approximately 4.7~K with SOC included.

Finally, we examine the influence of SOC on the superconducting behavior. 
The blue curves in Fig.~\ref{fig5}(b) show $\alpha^{2}F(\omega)$ and $\lambda(\omega)$ obtained without SOC. 
Compared with the SOC-included results, the most pronounced difference is the suppression of $\alpha^{2}F(\omega)$ in the energy window of 3.50--8.15~meV (shaded region in Fig.~\ref{fig5}(b)), which reduces the $\lambda$ from 0.83 to 0.71. 
To elucidate the microscopic origin of this suppression, we show in Fig.~\ref{fig5}(d) the difference of $\lambda_{\boldsymbol{q}}$ ($\Delta\lambda_{\boldsymbol{q}}$) calculated without and with SOC, contributed by the  phonons within this  energy window.
The  $\Delta\lambda_{\boldsymbol{q}}$ can be written as:
$$
 \Delta\lambda_{\boldsymbol{q}}(\omega_1,\omega_2) = \lambda^{\text{woSOC}}_{\boldsymbol{q}}(\omega_1,\omega_2) - \lambda^{\text{wSOC}}_{\boldsymbol{q}}(\omega_1,\omega_2).
$$
Fig.~\ref{fig5}(d) shows that without SOC, the $\boldsymbol{q}$ with the strongest coupling form an approximately sixfold-symmetric ring centered at $\Gamma$, with a representative vector marked by a vertical red double-headed arrow. 
As shown in Fig.~\ref{fig5}(g), the electronic states nested by this vector lie on the two hole pockets closest to $\Gamma$, similar to the SOC case described earlier.
Consequently, these two pockets exhibit relatively large $\Delta_{\boldsymbol{k}}$, while the remaining hole pockets show smaller $\Delta_{\boldsymbol{k}}$, giving rise to the anisotropy in the EPC and a broader $\Delta_{\boldsymbol{k}}$ distribution in energy without SOC (Fig.~\ref{fig5}(e)). 
When SOC is included, the dominant $\boldsymbol{q}$ in Fig.~\ref{fig5}(d) contract toward the $\Gamma$ point, and the ring-like distribution becomes significantly narrower. 
This behavior originates from the SOC-induced shrinking of the innermost hole pockets as described earlier, which reduces their contribution to $N(0)$ and suppresses inter-pocket scattering between the two hole pockets closest to $\Gamma$.
Consequently, the number of electronic states with large $\Delta_{\boldsymbol{k}}$ is markedly reduced, leading to a more concentrated $\Delta_{\boldsymbol{k}}$ distribution.
One see from Fig.~\ref{fig5}(e) that the $T_{\text{c}}$ is reduced from 7.1~K without SOC to about 4.7~K with SOC, further confirming the weakening of superconductivity by SOC.
Overall, these results demonstrate that superconductivity in the 2L is primarily driven by inter-pocket scattering between the two hole pockets closest to the $\Gamma$ point, and that SOC weakens this scattering by shrinking the inner pocket, thereby reducing the EPC strength.

We note that the above computational results for the 1L and 2L systems were obtained without spin polarization. 
As these systems were calculated to be nonmagnetic (see Fig.~S8 \citep{SM}), consistent with their bulk counterpart \citep{chakraborty2023observation,kindra2024magnetism,wang2020thickness}.

Given that alkali-metal intercalation has been predicted to induce  metal-to-superconductor transitions in transition metal compounds \citep{hong2024multigap,wu2021enhanced,huang2016dynamical,liao2025orbital}, we examine its role in the superconducting behavior of 2L CoTe$_2$.
We find that alkali-metal intercalation instead suppresses the superconductivity (see Fig.~S9 \citep{SM}). 
This suppression originates from the hole-pocket-dominated Fermi surface of the system. 
The electron doping via alkali-metal intercalation shrinks these hole pockets, thereby weakening the EPC and consequently suppressing superconductivity.

\section{Conclusion}

In summary, we have systematically investigated 1L and 2L CoTe$_2$ using first-principles calculations, focusing on their electronic structure, lattice dynamics, EPC, and superconductivity. 
Our key findings are as follows.
Firstly, 1L CoTe$_2$ is dynamically unstable at low temperatures. 
This instability manifests as extensive imaginary phonon modes near the M and K points, arising from collective Te$_{z}$ and Co$_{xy}$ vibrations. 
These unstable phonons exhibit strong EPC, originating from the combined effect of pronounced EPC matrix elements and Fermi-surface nesting. 
This interplay drives the inter-pocket scattering between $p$--$d$ hybridized states, which constitutes the primary mechanism for the dynamical instability.
Secondly, the transition from 1L to 2L stabilizes the crystal structure. 
The redistribution of Te‑$p_z$ electrons between the facing layers effectively electron-dopes and shrinks the original hole pockets, suppressing the inter-pocket scattering and reducing the EPC strength of the unstable modes. 
Concurrently, the formation of interlayer Te-Te quasi-bonds stiffens the Te$_{z}$ vibrations. 
These combined effects, introduced by the interlayer coupling, eliminate the imaginary phonons, converting them into low-energy modes with locally flattened dispersions. 
These modes mediate scattering between the two hole pockets closest to $\Gamma$, yielding superconductivity with an estimated $T_{\text{c}}\approx4.7$ K.
Finally, we analyze the role of SOC in 2L CoTe$_2$. 
SOC is found to shrink the innermost hole pocket, thereby weakening the EPC and superconductivity.
Our work demonstrates how interlayer coupling modulates structural stability and superconductivity in a two-dimensional material. 
These findings offer valuable insights for understanding and engineering quantum phases in layered transition-metal dichalcogenides, and present a system exhibiting distinct layer-dependent charge orders.

\begin{acknowledgements} 
   This work is supported by National Natural Science Foundation of China (12574264, 11804118), Guangdong Basic and Applied Basic Research Foundation (2025A1515010219, 2021A1515010041), the Science and Technology Planning Project of Guangzhou (202201010222), and the National Innovation and Entrepreneurship Training Program For Undergraduate in Jinan University (202410559005).
  The Calculations were performed on  high-performance computation cluster of Jinan University, and Tianhe Supercomputer System.
  Wenping Chen and Ziyun Zhang contributed equally to this work.
  \end{acknowledgements}
  

\bibliographystyle{apsrev4-2}

\begin{thebibliography}{54}%
\makeatletter
\providecommand \@ifxundefined [1]{%
 \@ifx{#1\undefined}
}%
\providecommand \@ifnum [1]{%
 \ifnum #1\expandafter \@firstoftwo
 \else \expandafter \@secondoftwo
 \fi
}%
\providecommand \@ifx [1]{%
 \ifx #1\expandafter \@firstoftwo
 \else \expandafter \@secondoftwo
 \fi
}%
\providecommand \natexlab [1]{#1}%
\providecommand \enquote  [1]{``#1''}%
\providecommand \bibnamefont  [1]{#1}%
\providecommand \bibfnamefont [1]{#1}%
\providecommand \citenamefont [1]{#1}%
\providecommand \href@noop [0]{\@secondoftwo}%
\providecommand \href [0]{\begingroup \@sanitize@url \@href}%
\providecommand \@href[1]{\@@startlink{#1}\@@href}%
\providecommand \@@href[1]{\endgroup#1\@@endlink}%
\providecommand \@sanitize@url [0]{\catcode `\\12\catcode `\$12\catcode
  `\&12\catcode `\#12\catcode `\^12\catcode `\_12\catcode `\%12\relax}%
\providecommand \@@startlink[1]{}%
\providecommand \@@endlink[0]{}%
\providecommand \url  [0]{\begingroup\@sanitize@url \@url }%
\providecommand \@url [1]{\endgroup\@href {#1}{\urlprefix }}%
\providecommand \urlprefix  [0]{URL }%
\providecommand \Eprint [0]{\href }%
\providecommand \doibase [0]{https://doi.org/}%
\providecommand \selectlanguage [0]{\@gobble}%
\providecommand \bibinfo  [0]{\@secondoftwo}%
\providecommand \bibfield  [0]{\@secondoftwo}%
\providecommand \translation [1]{[#1]}%
\providecommand \BibitemOpen [0]{}%
\providecommand \bibitemStop [0]{}%
\providecommand \bibitemNoStop [0]{.\EOS\space}%
\providecommand \EOS [0]{\spacefactor3000\relax}%
\providecommand \BibitemShut  [1]{\csname bibitem#1\endcsname}%
\let\auto@bib@innerbib\@empty
\bibitem [{\citenamefont {Ohtani}\ \emph {et~al.}(1984)\citenamefont {Ohtani},
  \citenamefont {Onoue},\ and\ \citenamefont {Nakahira}}]{ohtani1984phase}%
  \BibitemOpen
  \bibfield  {author} {\bibinfo {author} {\bibfnamefont {T.}~\bibnamefont
  {Ohtani}}, \bibinfo {author} {\bibfnamefont {S.}~\bibnamefont {Onoue}},\ and\
  \bibinfo {author} {\bibfnamefont {M.}~\bibnamefont {Nakahira}},\ }\href@noop
  {} {\bibfield  {journal} {\bibinfo  {journal} {Materials research bulletin}\
  }\textbf {\bibinfo {volume} {19}},\ \bibinfo {pages} {1367} (\bibinfo {year}
  {1984})}\BibitemShut {NoStop}%
\bibitem [{\citenamefont {Duvjir}\ \emph {et~al.}(2022)\citenamefont {Duvjir},
  \citenamefont {Jung}, \citenamefont {Ly}, \citenamefont {Lam}, \citenamefont
  {Chang}, \citenamefont {Lee}, \citenamefont {Kim},\ and\ \citenamefont
  {Kim}}]{duvjir2022fine}%
  \BibitemOpen
  \bibfield  {author} {\bibinfo {author} {\bibfnamefont {G.}~\bibnamefont
  {Duvjir}}, \bibinfo {author} {\bibfnamefont {J.-A.}\ \bibnamefont {Jung}},
  \bibinfo {author} {\bibfnamefont {T.~T.}\ \bibnamefont {Ly}}, \bibinfo
  {author} {\bibfnamefont {N.~H.}\ \bibnamefont {Lam}}, \bibinfo {author}
  {\bibfnamefont {Y.~J.}\ \bibnamefont {Chang}}, \bibinfo {author}
  {\bibfnamefont {S.}~\bibnamefont {Lee}}, \bibinfo {author} {\bibfnamefont
  {H.}~\bibnamefont {Kim}},\ and\ \bibinfo {author} {\bibfnamefont
  {J.}~\bibnamefont {Kim}},\ }\href@noop {} {\bibfield  {journal} {\bibinfo
  {journal} {APL Materials}\ }\textbf {\bibinfo {volume} {10}} (\bibinfo {year}
  {2022})}\BibitemShut {NoStop}%
\bibitem [{\citenamefont {Mitsuishi}\ \emph {et~al.}(2020)\citenamefont
  {Mitsuishi}, \citenamefont {Sugita}, \citenamefont {Bahramy}, \citenamefont
  {Kamitani}, \citenamefont {Sonobe}, \citenamefont {Sakano}, \citenamefont
  {Shimojima}, \citenamefont {Takahashi}, \citenamefont {Sakai}, \citenamefont
  {Horiba} \emph {et~al.}}]{mitsuishi2020switching}%
  \BibitemOpen
  \bibfield  {author} {\bibinfo {author} {\bibfnamefont {N.}~\bibnamefont
  {Mitsuishi}}, \bibinfo {author} {\bibfnamefont {Y.}~\bibnamefont {Sugita}},
  \bibinfo {author} {\bibfnamefont {M.}~\bibnamefont {Bahramy}}, \bibinfo
  {author} {\bibfnamefont {M.}~\bibnamefont {Kamitani}}, \bibinfo {author}
  {\bibfnamefont {T.}~\bibnamefont {Sonobe}}, \bibinfo {author} {\bibfnamefont
  {M.}~\bibnamefont {Sakano}}, \bibinfo {author} {\bibfnamefont
  {T.}~\bibnamefont {Shimojima}}, \bibinfo {author} {\bibfnamefont
  {H.}~\bibnamefont {Takahashi}}, \bibinfo {author} {\bibfnamefont
  {H.}~\bibnamefont {Sakai}}, \bibinfo {author} {\bibfnamefont
  {K.}~\bibnamefont {Horiba}}, \emph {et~al.},\ }\href@noop {} {\bibfield
  {journal} {\bibinfo  {journal} {Nature Communications}\ }\textbf {\bibinfo
  {volume} {11}},\ \bibinfo {pages} {2466} (\bibinfo {year}
  {2020})}\BibitemShut {NoStop}%
\bibitem [{\citenamefont {Chazarin}\ \emph {et~al.}(2024)\citenamefont
  {Chazarin}, \citenamefont {Lezoualc'h}, \citenamefont {Karn}, \citenamefont
  {Chou}, \citenamefont {Pai}, \citenamefont {Chacon}, \citenamefont {Girard},
  \citenamefont {Repain}, \citenamefont {Bellec}, \citenamefont {Rousset} \emph
  {et~al.}}]{chazarin2024spatially}%
  \BibitemOpen
  \bibfield  {author} {\bibinfo {author} {\bibfnamefont {U.}~\bibnamefont
  {Chazarin}}, \bibinfo {author} {\bibfnamefont {M.}~\bibnamefont
  {Lezoualc'h}}, \bibinfo {author} {\bibfnamefont {A.}~\bibnamefont {Karn}},
  \bibinfo {author} {\bibfnamefont {J.-P.}\ \bibnamefont {Chou}}, \bibinfo
  {author} {\bibfnamefont {W.~W.}\ \bibnamefont {Pai}}, \bibinfo {author}
  {\bibfnamefont {C.}~\bibnamefont {Chacon}}, \bibinfo {author} {\bibfnamefont
  {Y.}~\bibnamefont {Girard}}, \bibinfo {author} {\bibfnamefont
  {V.}~\bibnamefont {Repain}}, \bibinfo {author} {\bibfnamefont
  {A.}~\bibnamefont {Bellec}}, \bibinfo {author} {\bibfnamefont
  {S.}~\bibnamefont {Rousset}}, \emph {et~al.},\ }\href@noop {} {\bibfield
  {journal} {\bibinfo  {journal} {Nano Letters}\ }\textbf {\bibinfo {volume}
  {24}},\ \bibinfo {pages} {3470} (\bibinfo {year} {2024})}\BibitemShut
  {NoStop}%
\bibitem [{\citenamefont {Wu}\ \emph {et~al.}(2020)\citenamefont {Wu},
  \citenamefont {Wang}, \citenamefont {Guo}, \citenamefont {Yang},\ and\
  \citenamefont {Gao}}]{wu2020orbital}%
  \BibitemOpen
  \bibfield  {author} {\bibinfo {author} {\bibfnamefont {Q.}~\bibnamefont
  {Wu}}, \bibinfo {author} {\bibfnamefont {Z.}~\bibnamefont {Wang}}, \bibinfo
  {author} {\bibfnamefont {Y.}~\bibnamefont {Guo}}, \bibinfo {author}
  {\bibfnamefont {F.}~\bibnamefont {Yang}},\ and\ \bibinfo {author}
  {\bibfnamefont {C.}~\bibnamefont {Gao}},\ }\href@noop {} {\bibfield
  {journal} {\bibinfo  {journal} {Physical Review B}\ }\textbf {\bibinfo
  {volume} {101}},\ \bibinfo {pages} {205105} (\bibinfo {year}
  {2020})}\BibitemShut {NoStop}%
\bibitem [{\citenamefont {Liu}\ \emph {et~al.}(2020)\citenamefont {Liu},
  \citenamefont {Wu}, \citenamefont {Liu}, \citenamefont {Wang}, \citenamefont
  {Yao},\ and\ \citenamefont {Zhong}}]{liu2020multimorphism}%
  \BibitemOpen
  \bibfield  {author} {\bibinfo {author} {\bibfnamefont {M.}~\bibnamefont
  {Liu}}, \bibinfo {author} {\bibfnamefont {C.}~\bibnamefont {Wu}}, \bibinfo
  {author} {\bibfnamefont {Z.}~\bibnamefont {Liu}}, \bibinfo {author}
  {\bibfnamefont {Z.}~\bibnamefont {Wang}}, \bibinfo {author} {\bibfnamefont
  {D.-X.}\ \bibnamefont {Yao}},\ and\ \bibinfo {author} {\bibfnamefont
  {D.}~\bibnamefont {Zhong}},\ }\href@noop {} {\bibfield  {journal} {\bibinfo
  {journal} {Nano Research}\ }\textbf {\bibinfo {volume} {13}},\ \bibinfo
  {pages} {1733} (\bibinfo {year} {2020})}\BibitemShut {NoStop}%
\bibitem [{\citenamefont {Wang}\ \emph {et~al.}(2025)\citenamefont {Wang},
  \citenamefont {Wang},\ and\ \citenamefont {Lian}}]{wang2025unraveling}%
  \BibitemOpen
  \bibfield  {author} {\bibinfo {author} {\bibfnamefont {Q.}~\bibnamefont
  {Wang}}, \bibinfo {author} {\bibfnamefont {F.}~\bibnamefont {Wang}},\ and\
  \bibinfo {author} {\bibfnamefont {C.-S.}\ \bibnamefont {Lian}},\ }\href@noop
  {} {\bibfield  {journal} {\bibinfo  {journal} {Physical Review B}\ }\textbf
  {\bibinfo {volume} {112}},\ \bibinfo {pages} {245410} (\bibinfo {year}
  {2025})}\BibitemShut {NoStop}%
\bibitem [{\citenamefont {Lin}\ \emph {et~al.}(2020)\citenamefont {Lin},
  \citenamefont {Villaos}, \citenamefont {Hlevyack}, \citenamefont {Chen},
  \citenamefont {Liu}, \citenamefont {Hsu}, \citenamefont {Avila},
  \citenamefont {Mo}, \citenamefont {Chuang},\ and\ \citenamefont
  {Chiang}}]{lin2020dimensionality}%
  \BibitemOpen
  \bibfield  {author} {\bibinfo {author} {\bibfnamefont {M.-K.}\ \bibnamefont
  {Lin}}, \bibinfo {author} {\bibfnamefont {R.~A.~B.}\ \bibnamefont {Villaos}},
  \bibinfo {author} {\bibfnamefont {J.~A.}\ \bibnamefont {Hlevyack}}, \bibinfo
  {author} {\bibfnamefont {P.}~\bibnamefont {Chen}}, \bibinfo {author}
  {\bibfnamefont {R.-Y.}\ \bibnamefont {Liu}}, \bibinfo {author} {\bibfnamefont
  {C.-H.}\ \bibnamefont {Hsu}}, \bibinfo {author} {\bibfnamefont
  {J.}~\bibnamefont {Avila}}, \bibinfo {author} {\bibfnamefont {S.-K.}\
  \bibnamefont {Mo}}, \bibinfo {author} {\bibfnamefont {F.-C.}\ \bibnamefont
  {Chuang}},\ and\ \bibinfo {author} {\bibfnamefont {T.-C.}\ \bibnamefont
  {Chiang}},\ }\href@noop {} {\bibfield  {journal} {\bibinfo  {journal}
  {Physical Review Letters}\ }\textbf {\bibinfo {volume} {124}},\ \bibinfo
  {pages} {036402} (\bibinfo {year} {2020})}\BibitemShut {NoStop}%
\bibitem [{\citenamefont {Sun}\ \emph {et~al.}(2023)\citenamefont {Sun},
  \citenamefont {Deng}, \citenamefont {Wu}, \citenamefont {Hao}, \citenamefont
  {Ma}, \citenamefont {Ma}, \citenamefont {Zhao}, \citenamefont {Meng},
  \citenamefont {Ji}, \citenamefont {Ding} \emph {et~al.}}]{sun2023high}%
  \BibitemOpen
  \bibfield  {author} {\bibinfo {author} {\bibfnamefont {R.}~\bibnamefont
  {Sun}}, \bibinfo {author} {\bibfnamefont {J.}~\bibnamefont {Deng}}, \bibinfo
  {author} {\bibfnamefont {X.}~\bibnamefont {Wu}}, \bibinfo {author}
  {\bibfnamefont {M.}~\bibnamefont {Hao}}, \bibinfo {author} {\bibfnamefont
  {K.}~\bibnamefont {Ma}}, \bibinfo {author} {\bibfnamefont {Y.}~\bibnamefont
  {Ma}}, \bibinfo {author} {\bibfnamefont {C.}~\bibnamefont {Zhao}}, \bibinfo
  {author} {\bibfnamefont {D.}~\bibnamefont {Meng}}, \bibinfo {author}
  {\bibfnamefont {X.}~\bibnamefont {Ji}}, \bibinfo {author} {\bibfnamefont
  {Y.}~\bibnamefont {Ding}}, \emph {et~al.},\ }\href@noop {} {\bibfield
  {journal} {\bibinfo  {journal} {Nature Communications}\ }\textbf {\bibinfo
  {volume} {14}},\ \bibinfo {pages} {6689} (\bibinfo {year}
  {2023})}\BibitemShut {NoStop}%
\bibitem [{\citenamefont {Zhang}\ \emph {et~al.}(2022)\citenamefont {Zhang},
  \citenamefont {Rousuli}, \citenamefont {Zhang}, \citenamefont {Luo},
  \citenamefont {Guo}, \citenamefont {Cong}, \citenamefont {Lin}, \citenamefont
  {Bao}, \citenamefont {Zhang}, \citenamefont {Xu} \emph
  {et~al.}}]{zhang2022tailored}%
  \BibitemOpen
  \bibfield  {author} {\bibinfo {author} {\bibfnamefont {H.}~\bibnamefont
  {Zhang}}, \bibinfo {author} {\bibfnamefont {A.}~\bibnamefont {Rousuli}},
  \bibinfo {author} {\bibfnamefont {K.}~\bibnamefont {Zhang}}, \bibinfo
  {author} {\bibfnamefont {L.}~\bibnamefont {Luo}}, \bibinfo {author}
  {\bibfnamefont {C.}~\bibnamefont {Guo}}, \bibinfo {author} {\bibfnamefont
  {X.}~\bibnamefont {Cong}}, \bibinfo {author} {\bibfnamefont {Z.}~\bibnamefont
  {Lin}}, \bibinfo {author} {\bibfnamefont {C.}~\bibnamefont {Bao}}, \bibinfo
  {author} {\bibfnamefont {H.}~\bibnamefont {Zhang}}, \bibinfo {author}
  {\bibfnamefont {S.}~\bibnamefont {Xu}}, \emph {et~al.},\ }\href@noop {}
  {\bibfield  {journal} {\bibinfo  {journal} {Nature Physics}\ }\textbf
  {\bibinfo {volume} {18}},\ \bibinfo {pages} {1425} (\bibinfo {year}
  {2022})}\BibitemShut {NoStop}%
\bibitem [{\citenamefont {Slathia}\ \emph {et~al.}(2024)\citenamefont
  {Slathia}, \citenamefont {Wei}, \citenamefont {Tripathi}, \citenamefont
  {Tromer}, \citenamefont {Negedu}, \citenamefont {Boland}, \citenamefont
  {Sarkar}, \citenamefont {Galvao}, \citenamefont {Dalton},\ and\ \citenamefont
  {Tiwary}}]{slathia2024thickness}%
  \BibitemOpen
  \bibfield  {author} {\bibinfo {author} {\bibfnamefont {S.}~\bibnamefont
  {Slathia}}, \bibinfo {author} {\bibfnamefont {C.}~\bibnamefont {Wei}},
  \bibinfo {author} {\bibfnamefont {M.}~\bibnamefont {Tripathi}}, \bibinfo
  {author} {\bibfnamefont {R.}~\bibnamefont {Tromer}}, \bibinfo {author}
  {\bibfnamefont {S.~D.}\ \bibnamefont {Negedu}}, \bibinfo {author}
  {\bibfnamefont {C.~S.}\ \bibnamefont {Boland}}, \bibinfo {author}
  {\bibfnamefont {S.}~\bibnamefont {Sarkar}}, \bibinfo {author} {\bibfnamefont
  {D.~S.}\ \bibnamefont {Galvao}}, \bibinfo {author} {\bibfnamefont
  {A.}~\bibnamefont {Dalton}},\ and\ \bibinfo {author} {\bibfnamefont {C.~S.}\
  \bibnamefont {Tiwary}},\ }\href@noop {} {\bibfield  {journal} {\bibinfo
  {journal} {2D Materials}\ }\textbf {\bibinfo {volume} {11}},\ \bibinfo
  {pages} {035006} (\bibinfo {year} {2024})}\BibitemShut {NoStop}%
\bibitem [{\citenamefont {Siegfried}\ \emph {et~al.}(2023)\citenamefont
  {Siegfried}, \citenamefont {Bhandari}, \citenamefont {Qi}, \citenamefont
  {Ghimire}, \citenamefont {Joshi}, \citenamefont {Messegee}, \citenamefont
  {Beeson}, \citenamefont {Liu}, \citenamefont {Ghimire}, \citenamefont {Dang}
  \emph {et~al.}}]{siegfried2023cote2}%
  \BibitemOpen
  \bibfield  {author} {\bibinfo {author} {\bibfnamefont {P.~E.}\ \bibnamefont
  {Siegfried}}, \bibinfo {author} {\bibfnamefont {H.}~\bibnamefont {Bhandari}},
  \bibinfo {author} {\bibfnamefont {J.}~\bibnamefont {Qi}}, \bibinfo {author}
  {\bibfnamefont {R.}~\bibnamefont {Ghimire}}, \bibinfo {author} {\bibfnamefont
  {J.}~\bibnamefont {Joshi}}, \bibinfo {author} {\bibfnamefont {Z.~T.}\
  \bibnamefont {Messegee}}, \bibinfo {author} {\bibfnamefont {W.~B.}\
  \bibnamefont {Beeson}}, \bibinfo {author} {\bibfnamefont {K.}~\bibnamefont
  {Liu}}, \bibinfo {author} {\bibfnamefont {M.~P.}\ \bibnamefont {Ghimire}},
  \bibinfo {author} {\bibfnamefont {Y.}~\bibnamefont {Dang}}, \emph {et~al.},\
  }\href@noop {} {\bibfield  {journal} {\bibinfo  {journal} {Advanced
  Materials}\ }\textbf {\bibinfo {volume} {35}},\ \bibinfo {pages} {2300640}
  (\bibinfo {year} {2023})}\BibitemShut {NoStop}%
\bibitem [{\citenamefont {Chakraborty}\ \emph {et~al.}(2023)\citenamefont
  {Chakraborty}, \citenamefont {Fujii}, \citenamefont {Kuo}, \citenamefont
  {Lue}, \citenamefont {Politano}, \citenamefont {Vobornik},\ and\
  \citenamefont {Agarwal}}]{chakraborty2023observation}%
  \BibitemOpen
  \bibfield  {author} {\bibinfo {author} {\bibfnamefont {A.}~\bibnamefont
  {Chakraborty}}, \bibinfo {author} {\bibfnamefont {J.}~\bibnamefont {Fujii}},
  \bibinfo {author} {\bibfnamefont {C.-N.}\ \bibnamefont {Kuo}}, \bibinfo
  {author} {\bibfnamefont {C.~S.}\ \bibnamefont {Lue}}, \bibinfo {author}
  {\bibfnamefont {A.}~\bibnamefont {Politano}}, \bibinfo {author}
  {\bibfnamefont {I.}~\bibnamefont {Vobornik}},\ and\ \bibinfo {author}
  {\bibfnamefont {A.}~\bibnamefont {Agarwal}},\ }\href@noop {} {\bibfield
  {journal} {\bibinfo  {journal} {Physical Review B}\ }\textbf {\bibinfo
  {volume} {107}},\ \bibinfo {pages} {085406} (\bibinfo {year}
  {2023})}\BibitemShut {NoStop}%
\bibitem [{\citenamefont {Kindra}\ and\ \citenamefont
  {Singh}(2024)}]{kindra2024magnetism}%
  \BibitemOpen
  \bibfield  {author} {\bibinfo {author} {\bibfnamefont {K.}~\bibnamefont
  {Kindra}}\ and\ \bibinfo {author} {\bibfnamefont {R.}~\bibnamefont {Singh}},\
  }\href@noop {} {\bibfield  {journal} {\bibinfo  {journal} {Canadian Journal
  of Physics}\ }\textbf {\bibinfo {volume} {102}},\ \bibinfo {pages} {604}
  (\bibinfo {year} {2024})}\BibitemShut {NoStop}%
\bibitem [{\citenamefont {Wang}\ \emph {et~al.}(2020)\citenamefont {Wang},
  \citenamefont {Zhou}, \citenamefont {Zhang}, \citenamefont {Zhang},
  \citenamefont {Ma}, \citenamefont {Yang}, \citenamefont {Wang}, \citenamefont
  {Li}, \citenamefont {Meng}, \citenamefont {Jiang} \emph
  {et~al.}}]{wang2020thickness}%
  \BibitemOpen
  \bibfield  {author} {\bibinfo {author} {\bibfnamefont {X.}~\bibnamefont
  {Wang}}, \bibinfo {author} {\bibfnamefont {Z.}~\bibnamefont {Zhou}}, \bibinfo
  {author} {\bibfnamefont {P.}~\bibnamefont {Zhang}}, \bibinfo {author}
  {\bibfnamefont {S.}~\bibnamefont {Zhang}}, \bibinfo {author} {\bibfnamefont
  {Y.}~\bibnamefont {Ma}}, \bibinfo {author} {\bibfnamefont {W.}~\bibnamefont
  {Yang}}, \bibinfo {author} {\bibfnamefont {H.}~\bibnamefont {Wang}}, \bibinfo
  {author} {\bibfnamefont {B.}~\bibnamefont {Li}}, \bibinfo {author}
  {\bibfnamefont {L.}~\bibnamefont {Meng}}, \bibinfo {author} {\bibfnamefont
  {H.}~\bibnamefont {Jiang}}, \emph {et~al.},\ }\href@noop {} {\bibfield
  {journal} {\bibinfo  {journal} {Chemistry of Materials}\ }\textbf {\bibinfo
  {volume} {32}},\ \bibinfo {pages} {2321} (\bibinfo {year}
  {2020})}\BibitemShut {NoStop}%
\bibitem [{\citenamefont {Zhao}\ \emph {et~al.}(2020)\citenamefont {Zhao},
  \citenamefont {Song}, \citenamefont {Wang}, \citenamefont {Riis-Jensen},
  \citenamefont {Fu}, \citenamefont {Deng}, \citenamefont {Wan}, \citenamefont
  {Kang}, \citenamefont {Ning}, \citenamefont {Dan} \emph
  {et~al.}}]{zhao2020engineering}%
  \BibitemOpen
  \bibfield  {author} {\bibinfo {author} {\bibfnamefont {X.}~\bibnamefont
  {Zhao}}, \bibinfo {author} {\bibfnamefont {P.}~\bibnamefont {Song}}, \bibinfo
  {author} {\bibfnamefont {C.}~\bibnamefont {Wang}}, \bibinfo {author}
  {\bibfnamefont {A.~C.}\ \bibnamefont {Riis-Jensen}}, \bibinfo {author}
  {\bibfnamefont {W.}~\bibnamefont {Fu}}, \bibinfo {author} {\bibfnamefont
  {Y.}~\bibnamefont {Deng}}, \bibinfo {author} {\bibfnamefont {D.}~\bibnamefont
  {Wan}}, \bibinfo {author} {\bibfnamefont {L.}~\bibnamefont {Kang}}, \bibinfo
  {author} {\bibfnamefont {S.}~\bibnamefont {Ning}}, \bibinfo {author}
  {\bibfnamefont {J.}~\bibnamefont {Dan}}, \emph {et~al.},\ }\href@noop {}
  {\bibfield  {journal} {\bibinfo  {journal} {Nature}\ }\textbf {\bibinfo
  {volume} {581}},\ \bibinfo {pages} {171} (\bibinfo {year}
  {2020})}\BibitemShut {NoStop}%
\bibitem [{\citenamefont {Zhao}\ \emph {et~al.}(2024)\citenamefont {Zhao},
  \citenamefont {Liu}, \citenamefont {Zhou}, \citenamefont {Yang},\ and\
  \citenamefont {Jiang}}]{zhao2024two}%
  \BibitemOpen
  \bibfield  {author} {\bibinfo {author} {\bibfnamefont {Y.-Q.}\ \bibnamefont
  {Zhao}}, \bibinfo {author} {\bibfnamefont {Q.}~\bibnamefont {Liu}}, \bibinfo
  {author} {\bibfnamefont {B.-J.}\ \bibnamefont {Zhou}}, \bibinfo {author}
  {\bibfnamefont {G.}~\bibnamefont {Yang}},\ and\ \bibinfo {author}
  {\bibfnamefont {S.-L.}\ \bibnamefont {Jiang}},\ }\href@noop {} {\bibfield
  {journal} {\bibinfo  {journal} {Rare Metals}\ }\textbf {\bibinfo {volume}
  {43}},\ \bibinfo {pages} {5860} (\bibinfo {year} {2024})}\BibitemShut
  {NoStop}%
\bibitem [{\citenamefont {Quan}\ \emph {et~al.}(2024)\citenamefont {Quan},
  \citenamefont {Lu}, \citenamefont {Wu}, \citenamefont {Shang}, \citenamefont
  {Li}, \citenamefont {Hu}, \citenamefont {Wang}, \citenamefont {Zhang},
  \citenamefont {Zhou}, \citenamefont {Zhao} \emph
  {et~al.}}]{quan2024spontaneous}%
  \BibitemOpen
  \bibfield  {author} {\bibinfo {author} {\bibfnamefont {W.}~\bibnamefont
  {Quan}}, \bibinfo {author} {\bibfnamefont {Y.}~\bibnamefont {Lu}}, \bibinfo
  {author} {\bibfnamefont {Q.}~\bibnamefont {Wu}}, \bibinfo {author}
  {\bibfnamefont {C.}~\bibnamefont {Shang}}, \bibinfo {author} {\bibfnamefont
  {C.}~\bibnamefont {Li}}, \bibinfo {author} {\bibfnamefont {J.}~\bibnamefont
  {Hu}}, \bibinfo {author} {\bibfnamefont {J.}~\bibnamefont {Wang}}, \bibinfo
  {author} {\bibfnamefont {Z.}~\bibnamefont {Zhang}}, \bibinfo {author}
  {\bibfnamefont {S.}~\bibnamefont {Zhou}}, \bibinfo {author} {\bibfnamefont
  {J.}~\bibnamefont {Zhao}}, \emph {et~al.},\ }\href@noop {} {\bibfield
  {journal} {\bibinfo  {journal} {Advanced Functional Materials}\ }\textbf
  {\bibinfo {volume} {34}},\ \bibinfo {pages} {2315831} (\bibinfo {year}
  {2024})}\BibitemShut {NoStop}%
\bibitem [{\citenamefont {Wu}\ \emph {et~al.}(2024)\citenamefont {Wu},
  \citenamefont {Quan}, \citenamefont {Pan}, \citenamefont {Hu}, \citenamefont
  {Zhang}, \citenamefont {Wang}, \citenamefont {Zheng},\ and\ \citenamefont
  {Zhang}}]{wu2024atomically}%
  \BibitemOpen
  \bibfield  {author} {\bibinfo {author} {\bibfnamefont {Q.}~\bibnamefont
  {Wu}}, \bibinfo {author} {\bibfnamefont {W.}~\bibnamefont {Quan}}, \bibinfo
  {author} {\bibfnamefont {S.}~\bibnamefont {Pan}}, \bibinfo {author}
  {\bibfnamefont {J.}~\bibnamefont {Hu}}, \bibinfo {author} {\bibfnamefont
  {Z.}~\bibnamefont {Zhang}}, \bibinfo {author} {\bibfnamefont
  {J.}~\bibnamefont {Wang}}, \bibinfo {author} {\bibfnamefont {F.}~\bibnamefont
  {Zheng}},\ and\ \bibinfo {author} {\bibfnamefont {Y.}~\bibnamefont {Zhang}},\
  }\href@noop {} {\bibfield  {journal} {\bibinfo  {journal} {Nano Letters}\
  }\textbf {\bibinfo {volume} {24}},\ \bibinfo {pages} {7672} (\bibinfo {year}
  {2024})}\BibitemShut {NoStop}%
\bibitem [{\citenamefont {Liu}\ \emph {et~al.}(2024)\citenamefont {Liu},
  \citenamefont {Gong}, \citenamefont {Yin}, \citenamefont {Yi},\ and\
  \citenamefont {Liu}}]{liu2024emerging}%
  \BibitemOpen
  \bibfield  {author} {\bibinfo {author} {\bibfnamefont {Y.}~\bibnamefont
  {Liu}}, \bibinfo {author} {\bibfnamefont {Q.}~\bibnamefont {Gong}}, \bibinfo
  {author} {\bibfnamefont {Y.}~\bibnamefont {Yin}}, \bibinfo {author}
  {\bibfnamefont {M.}~\bibnamefont {Yi}},\ and\ \bibinfo {author}
  {\bibfnamefont {Y.}~\bibnamefont {Liu}},\ }\href@noop {} {\bibfield
  {journal} {\bibinfo  {journal} {Advanced Functional Materials}\ }\textbf
  {\bibinfo {volume} {34}},\ \bibinfo {pages} {2310372} (\bibinfo {year}
  {2024})}\BibitemShut {NoStop}%
\bibitem [{\citenamefont {Perdew}\ \emph {et~al.}(1996)\citenamefont {Perdew},
  \citenamefont {Burke},\ and\ \citenamefont {Ernzerhof}}]{pbe1}%
  \BibitemOpen
  \bibfield  {author} {\bibinfo {author} {\bibfnamefont {J.~P.}\ \bibnamefont
  {Perdew}}, \bibinfo {author} {\bibfnamefont {K.}~\bibnamefont {Burke}},\ and\
  \bibinfo {author} {\bibfnamefont {M.}~\bibnamefont {Ernzerhof}},\ }\href@noop
  {} {\bibfield  {journal} {\bibinfo  {journal} {Physical Review Letters}\
  }\textbf {\bibinfo {volume} {77}},\ \bibinfo {pages} {3865} (\bibinfo {year}
  {1996})}\BibitemShut {NoStop}%
\bibitem [{\citenamefont {Kresse}\ and\ \citenamefont {Joubert}(1999)}]{paw1}%
  \BibitemOpen
  \bibfield  {author} {\bibinfo {author} {\bibfnamefont {G.}~\bibnamefont
  {Kresse}}\ and\ \bibinfo {author} {\bibfnamefont {D.}~\bibnamefont
  {Joubert}},\ }\href@noop {} {\bibfield  {journal} {\bibinfo  {journal}
  {Physical Review B}\ }\textbf {\bibinfo {volume} {59}},\ \bibinfo {pages}
  {1758} (\bibinfo {year} {1999})}\BibitemShut {NoStop}%
\bibitem [{\citenamefont {Hamann}(2013)}]{oncv1}%
  \BibitemOpen
  \bibfield  {author} {\bibinfo {author} {\bibfnamefont {D.}~\bibnamefont
  {Hamann}},\ }\href@noop {} {\bibfield  {journal} {\bibinfo  {journal}
  {Physical Review B}\ }\textbf {\bibinfo {volume} {88}},\ \bibinfo {pages}
  {085117} (\bibinfo {year} {2013})}\BibitemShut {NoStop}%
\bibitem [{\citenamefont {Hamann}(2017)}]{oncv2}%
  \BibitemOpen
  \bibfield  {author} {\bibinfo {author} {\bibfnamefont {D.}~\bibnamefont
  {Hamann}},\ }\href@noop {} {\bibfield  {journal} {\bibinfo  {journal}
  {Physical Review B}\ }\textbf {\bibinfo {volume} {95}},\ \bibinfo {pages}
  {239906} (\bibinfo {year} {2017})}\BibitemShut {NoStop}%
\bibitem [{\citenamefont {Schlipf}\ and\ \citenamefont {Gygi}(2015)}]{oncv3}%
  \BibitemOpen
  \bibfield  {author} {\bibinfo {author} {\bibfnamefont {M.}~\bibnamefont
  {Schlipf}}\ and\ \bibinfo {author} {\bibfnamefont {F.}~\bibnamefont {Gygi}},\
  }\href@noop {} {\bibfield  {journal} {\bibinfo  {journal} {Computer Physics
  Communications}\ }\textbf {\bibinfo {volume} {196}},\ \bibinfo {pages} {36}
  (\bibinfo {year} {2015})}\BibitemShut {NoStop}%
\bibitem [{\citenamefont {Van~Setten}\ \emph {et~al.}(2018)\citenamefont
  {Van~Setten}, \citenamefont {Giantomassi}, \citenamefont {Bousquet},
  \citenamefont {Verstraete}, \citenamefont {Hamann}, \citenamefont {Gonze},\
  and\ \citenamefont {Rignanese}}]{dojo1}%
  \BibitemOpen
  \bibfield  {author} {\bibinfo {author} {\bibfnamefont {M.~J.}\ \bibnamefont
  {Van~Setten}}, \bibinfo {author} {\bibfnamefont {M.}~\bibnamefont
  {Giantomassi}}, \bibinfo {author} {\bibfnamefont {E.}~\bibnamefont
  {Bousquet}}, \bibinfo {author} {\bibfnamefont {M.~J.}\ \bibnamefont
  {Verstraete}}, \bibinfo {author} {\bibfnamefont {D.~R.}\ \bibnamefont
  {Hamann}}, \bibinfo {author} {\bibfnamefont {X.}~\bibnamefont {Gonze}},\ and\
  \bibinfo {author} {\bibfnamefont {G.-M.}\ \bibnamefont {Rignanese}},\
  }\href@noop {} {\bibfield  {journal} {\bibinfo  {journal} {Computer Physics
  Communications}\ }\textbf {\bibinfo {volume} {226}},\ \bibinfo {pages} {39}
  (\bibinfo {year} {2018})}\BibitemShut {NoStop}%
\bibitem [{\citenamefont {Grimme}\ \emph {et~al.}(2010)\citenamefont {Grimme},
  \citenamefont {Antony}, \citenamefont {Ehrlich},\ and\ \citenamefont
  {Krieg}}]{DFT-D3}%
  \BibitemOpen
  \bibfield  {author} {\bibinfo {author} {\bibfnamefont {S.}~\bibnamefont
  {Grimme}}, \bibinfo {author} {\bibfnamefont {J.}~\bibnamefont {Antony}},
  \bibinfo {author} {\bibfnamefont {S.}~\bibnamefont {Ehrlich}},\ and\ \bibinfo
  {author} {\bibfnamefont {H.}~\bibnamefont {Krieg}},\ }\href@noop {}
  {\bibfield  {journal} {\bibinfo  {journal} {The Journal of chemical physics}\
  }\textbf {\bibinfo {volume} {132}} (\bibinfo {year} {2010})}\BibitemShut
  {NoStop}%
\bibitem [{\citenamefont {Giannozzi}\ \emph {et~al.}(2009)\citenamefont
  {Giannozzi}, \citenamefont {Baroni}, \citenamefont {Bonini}, \citenamefont
  {Calandra}, \citenamefont {Car}, \citenamefont {Cavazzoni}, \citenamefont
  {Ceresoli}, \citenamefont {Chiarotti}, \citenamefont {Cococcioni},
  \citenamefont {Dabo} \emph {et~al.}}]{qe1}%
  \BibitemOpen
  \bibfield  {author} {\bibinfo {author} {\bibfnamefont {P.}~\bibnamefont
  {Giannozzi}}, \bibinfo {author} {\bibfnamefont {S.}~\bibnamefont {Baroni}},
  \bibinfo {author} {\bibfnamefont {N.}~\bibnamefont {Bonini}}, \bibinfo
  {author} {\bibfnamefont {M.}~\bibnamefont {Calandra}}, \bibinfo {author}
  {\bibfnamefont {R.}~\bibnamefont {Car}}, \bibinfo {author} {\bibfnamefont
  {C.}~\bibnamefont {Cavazzoni}}, \bibinfo {author} {\bibfnamefont
  {D.}~\bibnamefont {Ceresoli}}, \bibinfo {author} {\bibfnamefont {G.~L.}\
  \bibnamefont {Chiarotti}}, \bibinfo {author} {\bibfnamefont {M.}~\bibnamefont
  {Cococcioni}}, \bibinfo {author} {\bibfnamefont {I.}~\bibnamefont {Dabo}},
  \emph {et~al.},\ }\href@noop {} {\bibfield  {journal} {\bibinfo  {journal}
  {Journal of physics: Condensed matter}\ }\textbf {\bibinfo {volume} {21}},\
  \bibinfo {pages} {395502} (\bibinfo {year} {2009})}\BibitemShut {NoStop}%
\bibitem [{\citenamefont {Kresse}\ and\ \citenamefont
  {Furthm{\"u}ller}(1996)}]{vasp1}%
  \BibitemOpen
  \bibfield  {author} {\bibinfo {author} {\bibfnamefont {G.}~\bibnamefont
  {Kresse}}\ and\ \bibinfo {author} {\bibfnamefont {J.}~\bibnamefont
  {Furthm{\"u}ller}},\ }\href@noop {} {\bibfield  {journal} {\bibinfo
  {journal} {Physical Review B}\ }\textbf {\bibinfo {volume} {54}},\ \bibinfo
  {pages} {11169} (\bibinfo {year} {1996})}\BibitemShut {NoStop}%
\bibitem [{\citenamefont {Togo}\ \emph {et~al.}(2023)\citenamefont {Togo},
  \citenamefont {Chaput}, \citenamefont {Tadano},\ and\ \citenamefont
  {Tanaka}}]{phonopy1}%
  \BibitemOpen
  \bibfield  {author} {\bibinfo {author} {\bibfnamefont {A.}~\bibnamefont
  {Togo}}, \bibinfo {author} {\bibfnamefont {L.}~\bibnamefont {Chaput}},
  \bibinfo {author} {\bibfnamefont {T.}~\bibnamefont {Tadano}},\ and\ \bibinfo
  {author} {\bibfnamefont {I.}~\bibnamefont {Tanaka}},\ }\href@noop {}
  {\bibfield  {journal} {\bibinfo  {journal} {Journal of Physics: Condensed
  Matter}\ }\textbf {\bibinfo {volume} {35}},\ \bibinfo {pages} {353001}
  (\bibinfo {year} {2023})}\BibitemShut {NoStop}%
\bibitem [{\citenamefont {Togo}(2023)}]{phonopy2}%
  \BibitemOpen
  \bibfield  {author} {\bibinfo {author} {\bibfnamefont {A.}~\bibnamefont
  {Togo}},\ }\href@noop {} {\bibfield  {journal} {\bibinfo  {journal} {Journal
  of the Physical Society of Japan}\ }\textbf {\bibinfo {volume} {92}},\
  \bibinfo {pages} {012001} (\bibinfo {year} {2023})}\BibitemShut {NoStop}%
\bibitem [{\citenamefont {Momma}\ and\ \citenamefont {Izumi}(2011)}]{vesta1}%
  \BibitemOpen
  \bibfield  {author} {\bibinfo {author} {\bibfnamefont {K.}~\bibnamefont
  {Momma}}\ and\ \bibinfo {author} {\bibfnamefont {F.}~\bibnamefont {Izumi}},\
  }\href@noop {} {\bibfield  {journal} {\bibinfo  {journal} {Applied
  Crystallography}\ }\textbf {\bibinfo {volume} {44}},\ \bibinfo {pages} {1272}
  (\bibinfo {year} {2011})}\BibitemShut {NoStop}%
\bibitem [{\citenamefont {Ponc{\'e}}\ \emph {et~al.}(2016)\citenamefont
  {Ponc{\'e}}, \citenamefont {Margine}, \citenamefont {Verdi},\ and\
  \citenamefont {Giustino}}]{epw1}%
  \BibitemOpen
  \bibfield  {author} {\bibinfo {author} {\bibfnamefont {S.}~\bibnamefont
  {Ponc{\'e}}}, \bibinfo {author} {\bibfnamefont {E.~R.}\ \bibnamefont
  {Margine}}, \bibinfo {author} {\bibfnamefont {C.}~\bibnamefont {Verdi}},\
  and\ \bibinfo {author} {\bibfnamefont {F.}~\bibnamefont {Giustino}},\
  }\href@noop {} {\bibfield  {journal} {\bibinfo  {journal} {Computer Physics
  Communications}\ }\textbf {\bibinfo {volume} {209}},\ \bibinfo {pages} {116}
  (\bibinfo {year} {2016})}\BibitemShut {NoStop}%
\bibitem [{\citenamefont {Mostofi}\ \emph {et~al.}(2008)\citenamefont
  {Mostofi}, \citenamefont {Yates}, \citenamefont {Lee}, \citenamefont {Souza},
  \citenamefont {Vanderbilt},\ and\ \citenamefont {Marzari}}]{w901}%
  \BibitemOpen
  \bibfield  {author} {\bibinfo {author} {\bibfnamefont {A.~A.}\ \bibnamefont
  {Mostofi}}, \bibinfo {author} {\bibfnamefont {J.~R.}\ \bibnamefont {Yates}},
  \bibinfo {author} {\bibfnamefont {Y.-S.}\ \bibnamefont {Lee}}, \bibinfo
  {author} {\bibfnamefont {I.}~\bibnamefont {Souza}}, \bibinfo {author}
  {\bibfnamefont {D.}~\bibnamefont {Vanderbilt}},\ and\ \bibinfo {author}
  {\bibfnamefont {N.}~\bibnamefont {Marzari}},\ }\href@noop {} {\bibfield
  {journal} {\bibinfo  {journal} {Computer Physics Communications}\ }\textbf
  {\bibinfo {volume} {178}},\ \bibinfo {pages} {685} (\bibinfo {year}
  {2008})}\BibitemShut {NoStop}%
\bibitem [{\citenamefont {Margine}\ and\ \citenamefont
  {Giustino}(2013)}]{anisotropicEPC1}%
  \BibitemOpen
  \bibfield  {author} {\bibinfo {author} {\bibfnamefont {E.~R.}\ \bibnamefont
  {Margine}}\ and\ \bibinfo {author} {\bibfnamefont {F.}~\bibnamefont
  {Giustino}},\ }\href@noop {} {\bibfield  {journal} {\bibinfo  {journal}
  {Physical Review B}\ }\textbf {\bibinfo {volume} {87}},\ \bibinfo {pages}
  {024505} (\bibinfo {year} {2013})}\BibitemShut {NoStop}%
\bibitem [{\citenamefont {Errea}\ \emph {et~al.}(2014)\citenamefont {Errea},
  \citenamefont {Calandra},\ and\ \citenamefont {Mauri}}]{sscha1}%
  \BibitemOpen
  \bibfield  {author} {\bibinfo {author} {\bibfnamefont {I.}~\bibnamefont
  {Errea}}, \bibinfo {author} {\bibfnamefont {M.}~\bibnamefont {Calandra}},\
  and\ \bibinfo {author} {\bibfnamefont {F.}~\bibnamefont {Mauri}},\
  }\href@noop {} {\bibfield  {journal} {\bibinfo  {journal} {Physical Review
  B}\ }\textbf {\bibinfo {volume} {89}},\ \bibinfo {pages} {064302} (\bibinfo
  {year} {2014})}\BibitemShut {NoStop}%
\bibitem [{\citenamefont {Slimani}\ \emph {et~al.}(2017)\citenamefont
  {Slimani}, \citenamefont {Khemakhem},\ and\ \citenamefont
  {Boukheddaden}}]{sscha2}%
  \BibitemOpen
  \bibfield  {author} {\bibinfo {author} {\bibfnamefont {A.}~\bibnamefont
  {Slimani}}, \bibinfo {author} {\bibfnamefont {H.}~\bibnamefont {Khemakhem}},\
  and\ \bibinfo {author} {\bibfnamefont {K.}~\bibnamefont {Boukheddaden}},\
  }\href@noop {} {\bibfield  {journal} {\bibinfo  {journal} {Physical Review
  B}\ }\textbf {\bibinfo {volume} {95}},\ \bibinfo {pages} {174104} (\bibinfo
  {year} {2017})}\BibitemShut {NoStop}%
\bibitem [{\citenamefont {Monacelli}\ \emph {et~al.}(2021)\citenamefont
  {Monacelli}, \citenamefont {Bianco}, \citenamefont {Cherubini}, \citenamefont
  {Calandra}, \citenamefont {Errea},\ and\ \citenamefont {Mauri}}]{sscha3}%
  \BibitemOpen
  \bibfield  {author} {\bibinfo {author} {\bibfnamefont {L.}~\bibnamefont
  {Monacelli}}, \bibinfo {author} {\bibfnamefont {R.}~\bibnamefont {Bianco}},
  \bibinfo {author} {\bibfnamefont {M.}~\bibnamefont {Cherubini}}, \bibinfo
  {author} {\bibfnamefont {M.}~\bibnamefont {Calandra}}, \bibinfo {author}
  {\bibfnamefont {I.}~\bibnamefont {Errea}},\ and\ \bibinfo {author}
  {\bibfnamefont {F.}~\bibnamefont {Mauri}},\ }\href@noop {} {\bibfield
  {journal} {\bibinfo  {journal} {Journal of Physics: Condensed Matter}\
  }\textbf {\bibinfo {volume} {33}},\ \bibinfo {pages} {363001} (\bibinfo
  {year} {2021})}\BibitemShut {NoStop}%
\bibitem [{\citenamefont {Zhang}\ \emph {et~al.}(2018)\citenamefont {Zhang},
  \citenamefont {Han}, \citenamefont {Wang}, \citenamefont {Car},\ and\
  \citenamefont {E}}]{dpmd1}%
  \BibitemOpen
  \bibfield  {author} {\bibinfo {author} {\bibfnamefont {L.}~\bibnamefont
  {Zhang}}, \bibinfo {author} {\bibfnamefont {J.}~\bibnamefont {Han}}, \bibinfo
  {author} {\bibfnamefont {H.}~\bibnamefont {Wang}}, \bibinfo {author}
  {\bibfnamefont {R.}~\bibnamefont {Car}},\ and\ \bibinfo {author}
  {\bibfnamefont {W.}~\bibnamefont {E}},\ }\href@noop {} {\bibfield  {journal}
  {\bibinfo  {journal} {Physical Review Letters}\ }\textbf {\bibinfo {volume}
  {120}},\ \bibinfo {pages} {143001} (\bibinfo {year} {2018})}\BibitemShut
  {NoStop}%
\bibitem [{\citenamefont {Wang}\ \emph {et~al.}(2018)\citenamefont {Wang},
  \citenamefont {Zhang}, \citenamefont {Han} \emph {et~al.}}]{dpmd2}%
  \BibitemOpen
  \bibfield  {author} {\bibinfo {author} {\bibfnamefont {H.}~\bibnamefont
  {Wang}}, \bibinfo {author} {\bibfnamefont {L.}~\bibnamefont {Zhang}},
  \bibinfo {author} {\bibfnamefont {J.}~\bibnamefont {Han}}, \emph {et~al.},\
  }\href@noop {} {\bibfield  {journal} {\bibinfo  {journal} {Computer Physics
  Communications}\ }\textbf {\bibinfo {volume} {228}},\ \bibinfo {pages} {178}
  (\bibinfo {year} {2018})}\BibitemShut {NoStop}%
\bibitem [{SM()}]{SM}%
  \BibitemOpen
  \href@noop {} {\bibinfo  {journal} {Supplemental Material}\ }\BibitemShut
  {NoStop}%
\bibitem [{\citenamefont {Zheng}\ \emph {et~al.}(2020)\citenamefont {Zheng},
  \citenamefont {Li}, \citenamefont {Tan}, \citenamefont {Lin}, \citenamefont
  {Xiong}, \citenamefont {Chen},\ and\ \citenamefont
  {Feng}}]{zheng2020emergent}%
  \BibitemOpen
\bibfield  {journal} {  }\bibfield  {author} {\bibinfo {author} {\bibfnamefont
  {F.}~\bibnamefont {Zheng}}, \bibinfo {author} {\bibfnamefont {X.-B.}\
  \bibnamefont {Li}}, \bibinfo {author} {\bibfnamefont {P.}~\bibnamefont
  {Tan}}, \bibinfo {author} {\bibfnamefont {Y.}~\bibnamefont {Lin}}, \bibinfo
  {author} {\bibfnamefont {L.}~\bibnamefont {Xiong}}, \bibinfo {author}
  {\bibfnamefont {X.}~\bibnamefont {Chen}},\ and\ \bibinfo {author}
  {\bibfnamefont {J.}~\bibnamefont {Feng}},\ }\href@noop {} {\bibfield
  {journal} {\bibinfo  {journal} {Physical Review B}\ }\textbf {\bibinfo
  {volume} {101}},\ \bibinfo {pages} {100505} (\bibinfo {year}
  {2020})}\BibitemShut {NoStop}%
\bibitem [{\citenamefont {Yan}\ \emph {et~al.}(2025)\citenamefont {Yan},
  \citenamefont {Xiong},\ and\ \citenamefont {Zheng}}]{yan2025charge}%
  \BibitemOpen
  \bibfield  {author} {\bibinfo {author} {\bibfnamefont {Y.}~\bibnamefont
  {Yan}}, \bibinfo {author} {\bibfnamefont {L.}~\bibnamefont {Xiong}},\ and\
  \bibinfo {author} {\bibfnamefont {F.}~\bibnamefont {Zheng}},\ }\href@noop {}
  {\bibfield  {journal} {\bibinfo  {journal} {Physical Review B}\ }\textbf
  {\bibinfo {volume} {111}},\ \bibinfo {pages} {205437} (\bibinfo {year}
  {2025})}\BibitemShut {NoStop}%
\bibitem [{\citenamefont {Wang}\ \emph {et~al.}(2023)\citenamefont {Wang},
  \citenamefont {Chen}, \citenamefont {Mo}, \citenamefont {Zhou}, \citenamefont
  {Loh},\ and\ \citenamefont {Feng}}]{wang2023decisive}%
  \BibitemOpen
  \bibfield  {author} {\bibinfo {author} {\bibfnamefont {Z.}~\bibnamefont
  {Wang}}, \bibinfo {author} {\bibfnamefont {C.}~\bibnamefont {Chen}}, \bibinfo
  {author} {\bibfnamefont {J.}~\bibnamefont {Mo}}, \bibinfo {author}
  {\bibfnamefont {J.}~\bibnamefont {Zhou}}, \bibinfo {author} {\bibfnamefont
  {K.~P.}\ \bibnamefont {Loh}},\ and\ \bibinfo {author} {\bibfnamefont {Y.~P.}\
  \bibnamefont {Feng}},\ }\href@noop {} {\bibfield  {journal} {\bibinfo
  {journal} {Physical Review Research}\ }\textbf {\bibinfo {volume} {5}},\
  \bibinfo {pages} {013218} (\bibinfo {year} {2023})}\BibitemShut {NoStop}%
\bibitem [{\citenamefont {Luo}\ \emph {et~al.}(2023)\citenamefont {Luo},
  \citenamefont {Zhang}, \citenamefont {Shi},\ and\ \citenamefont
  {Zheng}}]{luo2023emergent}%
  \BibitemOpen
  \bibfield  {author} {\bibinfo {author} {\bibfnamefont {T.}~\bibnamefont
  {Luo}}, \bibinfo {author} {\bibfnamefont {M.}~\bibnamefont {Zhang}}, \bibinfo
  {author} {\bibfnamefont {J.}~\bibnamefont {Shi}},\ and\ \bibinfo {author}
  {\bibfnamefont {F.}~\bibnamefont {Zheng}},\ }\href@noop {} {\bibfield
  {journal} {\bibinfo  {journal} {Physical Review B}\ }\textbf {\bibinfo
  {volume} {107}},\ \bibinfo {pages} {L161401} (\bibinfo {year}
  {2023})}\BibitemShut {NoStop}%
\bibitem [{\citenamefont {Giustino}(2017)}]{giustino2017electron}%
  \BibitemOpen
  \bibfield  {author} {\bibinfo {author} {\bibfnamefont {F.}~\bibnamefont
  {Giustino}},\ }\href@noop {} {\bibfield  {journal} {\bibinfo  {journal}
  {Reviews of Modern Physics}\ }\textbf {\bibinfo {volume} {89}},\ \bibinfo
  {pages} {015003} (\bibinfo {year} {2017})}\BibitemShut {NoStop}%
\bibitem [{\citenamefont {Felix~Flicker}(2016)}]{felix2016}%
  \BibitemOpen
  \bibfield  {author} {\bibinfo {author} {\bibfnamefont {J.~v.~W.}\
  \bibnamefont {Felix~Flicker}},\ }\href@noop {} {\bibfield  {journal}
  {\bibinfo  {journal} {Physical Review B}\ }\textbf {\bibinfo {volume} {94}},\
  \bibinfo {pages} {235135} (\bibinfo {year} {2016})}\BibitemShut {NoStop}%
\bibitem [{\citenamefont {Zhu}\ \emph {et~al.}(2015)\citenamefont {Zhu},
  \citenamefont {Cao}, \citenamefont {Zhang}, \citenamefont {Plummer},\ and\
  \citenamefont {Guo}}]{zhu2015classification}%
  \BibitemOpen
  \bibfield  {author} {\bibinfo {author} {\bibfnamefont {X.}~\bibnamefont
  {Zhu}}, \bibinfo {author} {\bibfnamefont {Y.}~\bibnamefont {Cao}}, \bibinfo
  {author} {\bibfnamefont {J.}~\bibnamefont {Zhang}}, \bibinfo {author}
  {\bibfnamefont {E.}~\bibnamefont {Plummer}},\ and\ \bibinfo {author}
  {\bibfnamefont {J.}~\bibnamefont {Guo}},\ }\href@noop {} {\bibfield
  {journal} {\bibinfo  {journal} {Proceedings of the National Academy of
  Sciences}\ }\textbf {\bibinfo {volume} {112}},\ \bibinfo {pages} {2367}
  (\bibinfo {year} {2015})}\BibitemShut {NoStop}%
\bibitem [{\citenamefont {Zheng}\ and\ \citenamefont
  {Feng}(2019)}]{zheng2019electron}%
  \BibitemOpen
  \bibfield  {author} {\bibinfo {author} {\bibfnamefont {F.}~\bibnamefont
  {Zheng}}\ and\ \bibinfo {author} {\bibfnamefont {J.}~\bibnamefont {Feng}},\
  }\href@noop {} {\bibfield  {journal} {\bibinfo  {journal} {Physical Review
  B}\ }\textbf {\bibinfo {volume} {99}},\ \bibinfo {pages} {161119} (\bibinfo
  {year} {2019})}\BibitemShut {NoStop}%
\bibitem [{\citenamefont {Lifshitz}\ \emph {et~al.}(1960)\citenamefont
  {Lifshitz} \emph {et~al.}}]{lifshitz1960anomalies}%
  \BibitemOpen
  \bibfield  {author} {\bibinfo {author} {\bibfnamefont {I.}~\bibnamefont
  {Lifshitz}} \emph {et~al.},\ }\href@noop {} {\bibfield  {journal} {\bibinfo
  {journal} {Sov. Phys. JETP}\ }\textbf {\bibinfo {volume} {11}},\ \bibinfo
  {pages} {1130} (\bibinfo {year} {1960})}\BibitemShut {NoStop}%
\bibitem [{\citenamefont {Hong}\ \emph {et~al.}(2024)\citenamefont {Hong},
  \citenamefont {Wu}, \citenamefont {Li},\ and\ \citenamefont
  {Zheng}}]{hong2024multigap}%
  \BibitemOpen
  \bibfield  {author} {\bibinfo {author} {\bibfnamefont {C.}~\bibnamefont
  {Hong}}, \bibinfo {author} {\bibfnamefont {D.}~\bibnamefont {Wu}}, \bibinfo
  {author} {\bibfnamefont {X.-B.}\ \bibnamefont {Li}},\ and\ \bibinfo {author}
  {\bibfnamefont {F.}~\bibnamefont {Zheng}},\ }\href@noop {} {\bibfield
  {journal} {\bibinfo  {journal} {Physical Review B}\ }\textbf {\bibinfo
  {volume} {109}},\ \bibinfo {pages} {064515} (\bibinfo {year}
  {2024})}\BibitemShut {NoStop}%
\bibitem [{\citenamefont {Wu}\ \emph {et~al.}(2021)\citenamefont {Wu},
  \citenamefont {Lin}, \citenamefont {Xiong}, \citenamefont {Li}, \citenamefont
  {Luo}, \citenamefont {Chen},\ and\ \citenamefont {Zheng}}]{wu2021enhanced}%
  \BibitemOpen
  \bibfield  {author} {\bibinfo {author} {\bibfnamefont {D.}~\bibnamefont
  {Wu}}, \bibinfo {author} {\bibfnamefont {Y.}~\bibnamefont {Lin}}, \bibinfo
  {author} {\bibfnamefont {L.}~\bibnamefont {Xiong}}, \bibinfo {author}
  {\bibfnamefont {J.}~\bibnamefont {Li}}, \bibinfo {author} {\bibfnamefont
  {T.}~\bibnamefont {Luo}}, \bibinfo {author} {\bibfnamefont {D.}~\bibnamefont
  {Chen}},\ and\ \bibinfo {author} {\bibfnamefont {F.}~\bibnamefont {Zheng}},\
  }\href@noop {} {\bibfield  {journal} {\bibinfo  {journal} {Physical Review
  B}\ }\textbf {\bibinfo {volume} {103}},\ \bibinfo {pages} {224502} (\bibinfo
  {year} {2021})}\BibitemShut {NoStop}%
\bibitem [{\citenamefont {Huang}\ \emph {et~al.}(2016)\citenamefont {Huang},
  \citenamefont {Xing},\ and\ \citenamefont {Xing}}]{huang2016dynamical}%
  \BibitemOpen
  \bibfield  {author} {\bibinfo {author} {\bibfnamefont {G.}~\bibnamefont
  {Huang}}, \bibinfo {author} {\bibfnamefont {Z.}~\bibnamefont {Xing}},\ and\
  \bibinfo {author} {\bibfnamefont {D.}~\bibnamefont {Xing}},\ }\href@noop {}
  {\bibfield  {journal} {\bibinfo  {journal} {Physical Review B}\ }\textbf
  {\bibinfo {volume} {93}},\ \bibinfo {pages} {104511} (\bibinfo {year}
  {2016})}\BibitemShut {NoStop}%
\bibitem [{\citenamefont {Liao}\ \emph {et~al.}(2025)\citenamefont {Liao},
  \citenamefont {Liu}, \citenamefont {Zhang}, \citenamefont {Zhao},
  \citenamefont {Wang}, \citenamefont {Wang}, \citenamefont {Jia},\ and\
  \citenamefont {Cho}}]{liao2025orbital}%
  \BibitemOpen
  \bibfield  {author} {\bibinfo {author} {\bibfnamefont {X.}~\bibnamefont
  {Liao}}, \bibinfo {author} {\bibfnamefont {L.}~\bibnamefont {Liu}}, \bibinfo
  {author} {\bibfnamefont {L.}~\bibnamefont {Zhang}}, \bibinfo {author}
  {\bibfnamefont {X.}~\bibnamefont {Zhao}}, \bibinfo {author} {\bibfnamefont
  {C.}~\bibnamefont {Wang}}, \bibinfo {author} {\bibfnamefont {B.}~\bibnamefont
  {Wang}}, \bibinfo {author} {\bibfnamefont {Y.}~\bibnamefont {Jia}},\ and\
  \bibinfo {author} {\bibfnamefont {J.}~\bibnamefont {Cho}},\ }\href@noop {}
  {\bibfield  {journal} {\bibinfo  {journal} {Physical Review B}\ }\textbf
  {\bibinfo {volume} {111}},\ \bibinfo {pages} {L020505} (\bibinfo {year}
  {2025})}\BibitemShut {NoStop}%
\end{thebibliography}

%

\end{document}